\documentstyle[aaspp4,epsfig]{article}
 \def\simlt{\lower.5ex\hbox{$\; \buildrel < \over \sim \;$}}
  \def\simgt{\lower.5ex\hbox{$\; \buildrel > \over \sim \;$}}
\def\mag{\mbox{ mag}}
\def\kms{\mbox{ km s$^{-1}$}}
\def\mpc{\mbox{ Mpc}}
\def\kpc{\mbox{ kpc}}
\def\pc{\mbox{ pc}}

\def\yr{\mbox{ yr}}
\def\msun{\mbox{ M}_\odot}
\def\micron{$\mu$m~}

\def\oiii{[O\,{\scriptsize III}]~}
\def\oiiilam{[O\,{\scriptsize III}]\,$\lambda\lambda4959,5007$}
\def\feii{Fe\,{\scriptsize II}~}
\def\hb{H$\beta$ }
\def\dmr{\mbox{DMR}}

\begin{document}

\title{Spectroscopic Gravitational Lensing and Limits on the Dark
  Matter Substructure in Q2237+0305}

\author{R. Benton Metcalf\footnote{Hubble Fellow}}
\affil{\it Department of Astronomy and Astrophysics, University of California,
Santa Cruz, CA 95064 USA}
\and
\author{Leonidas A. Moustakas}
\affil{\it Space Telescope Science Institute, 3700 San Martin Drive,
  Baltimore, MD 21218 USA}
\and
\author{Andrew J. Bunker}
\affil{\it Institute of Astronomy, Cambridge University, Madingley Road,
  CB3 0HA United Kingdom}
\and
\author{Ian R. Parry}
\affil{\it Institute of Astronomy, Cambridge University, Madingley Road,
  CB3 0HA United Kingdom}

\abstract{Spatially resolved spectroscopic 
data from the CIRPASS integral field unit (IFU) on Gemini are used to
measure the 
gravitational lensing of the 4--image quasar Q2237+0305 on different
size scales. A method for measuring the substructure present in the lens
using observations at multiple wavelengths is demonstrated to be very
effective and independent of many of the degeneracies inherent in
previous methods. 
The magnification ratios of the QSO's narrow line region (NLR) and broad
line region (BLR) are measured and found to be disagree with each other
and with the published radio and mid-infrared magnification ratios.  The
disagreement between the BLR ratios and the radio/mid-infrared ratios is
interpreted as microlensing by stars in the lens galaxy of the 
BLR.  This implies that the mid-infrared emission region is larger than the
BLR and the BLR is $\simlt 0.1\pc$ in size.  We find a small difference in
the shape of the H$\beta$ line in image~A when compared to the other
images.  We consider this difference too small and symmetric to be
considered strong evidence for rotation or large scale infall in the
H$\beta$ emission region.  The disagreement between the
radio/mid-infrared ratios and the NLR ratios is interpreted as a
signature of substructure on a larger scale, possibly the missing small
scale structure predicted by the standard cold dark matter (CDM) model. 
Extensive lensing simulations are performed to obtain a lower limit on
the amount of substructure that is required to cause this discrepancy as
a function of its mass and the radial profile of the host lens.  
The substructure surface density is degenerate with the radial profile
of the host lens, but if the expectations of the CDM model are taken
into account certain radial profiles and substructure surface densities
can be ruled out.  A substructure mass scale as large as $10^8\msun$ is
strongly disfavored while $10^4\msun$ is too small if the radio 
and mid-infrared emission regions have the expected sizes of $\sim
10\pc$.  The standard elliptical isothermal lens mass profile is not
compatible with a substructure surface density of $\Sigma_{\rm sub}< 280\msun\pc^{-2}$ at the 95\% confidence level.  This is $4 - 7\%$
of the galaxy's surface density (depending on which image position is
used to evaluate this).  The required 
substructure surface density at the required mass scale is high in
comparison with the present expectations within the CDM model.  Lens mass
profiles that are flatter than isothermal -- where the surface density
in dark matter is higher at the image positions -- are compatible with
smaller quantities of substructure.
}

\section{Introduction}

In this paper we present observations of the lensed quasi-stellar object
(QSO) Q2237+0305 and use them to address fundamental questions concerning
the Cold Dark Matter (CDM) cosmological model and in addition to
investigate the internal structure of the QSO. 
The CDM cosmological model has recently experienced
great successes in predicting the distribution of galaxies and the
fluctuations in the cosmic microwave background.  However, there remain some
challenges to this model on the scale of galaxies themselves.
Simulations of structure formation within this model predict that a high
level of substructure will survive tidal disruption within the large
dark matter (DM) halos that surround galaxies.
N-body simulations indicate that there should be plentiful subhalos in a
$L_*$ galaxy's halo, with masses above $10^7\msun$ -- smaller mass scales
are presently inaccessible due to limitations in resolution.
If these subhalos contained stars, there would be significantly more
dwarf galaxies in the Local Group than are seen \markcite{1999MNRAS.310.1147M,
2001ApJ...554..903K}({Moore} {et~al.} 1999; {Klypin} {et~al.} 2001).  

Various solutions to this ``dwarf satellite problem'' have been
proposed.  In particular, if  the dark matter is warm (WDM) rather than
cold then the  small scale structure is reduced while the
large scale (successful) properties of the standard paradigm are
preserved.  Another explanation for the absence of observable substructure is that the star
formation in these  small DM halos was  inhibited by
ionizing radiation at an early  stage in the universe
(e.g. \markcite{astro-ph/0107507,2000ApJ...539..517B,Dong2003}{Somerville} (2002); {Bullock}, {Kravtsov}, \&  {Weinberg} (2000); Dong, Murray, \& Lin (2003)).  In this case the
substructure does exist, but is invisible.  \markcite{2001ApJ...563....9M}{Metcalf} \& {Madau} (2001)
showed that this substructure could be detected in gravitational lenses.
There is already some
evidence in support of the need for substructure to explain the
relative magnifications in strong lenses; disentangling its signature
from other (also interesting) effects is the goal of this paper.

The extraction of useful information from QSO lens systems has long been
hampered by uncertainties and ambiguities related to modeling the lens'
mass distribution.  This has been the case for estimates of the Hubble
parameter from lensing and it has necessitated long term monitoring
in the optical to detect microlensing.  Recently it has become an
important issue in testing the CDM cosmological model
by detecting the dark substructure predicted in this model. 
From the positions of the images and lens galaxy of a 4-image lens
system, a parametric smooth mass model for the lensing galaxy and halo
can be computed (including external shear), along with predictions for
the three independent flux ratios.  
Measuring deviations between the
predictions and the (differential-reddening-corrected) observed fluxes
in real lenses has been used to argue that substructure is present in
almost all QSO lens
systems \markcite{2001ApJ...563....9M,2002ApJ...565...17C,2002ApJ...567L...5M,Dalal2002}({Metcalf} \& {Madau} 2001; {Chiba} 2002; {Metcalf} \& {Zhao} 2002; {Dalal} \& {Kochanek} 2002).
 Some substructure has long been considered a possible explanation for
inconsistencies in the observed flux ratio of B1422+231.
\markcite{1998MNRAS.295..587M}{Mao} \& {Schneider} (1998) argued that in this case the source quasar is
near a cusp caustic in the lensing map and because of this the sum of the
magnifications of the three closest images should be close to zero for
any smooth lens model.  B1422+231 and all other known cusp caustic
lenses violate this relation.
It is unclear how far this argument can be taken since the
proximity of the source to the caustic depends on the lens model itself
\markcite{2002ApJ...580..468G}(see {Gaudi} \& {Petters} 2002) and stellar disks may cause the
relation to be violated even very close to the cusp
\markcite{astro-ph/0306238}(Brada{\v c} {et~al.} 2003).  It is also unclear what mass or size the
substructure needs to be for this argument to break down, i.e. how to
measure properties of the substructure from violations of these
relations.  In any case, most lens systems do not have such 
advantageously located source quasars and the predicted flux ratios can be
strongly dependent on the parametric lens models used while the image
positions are degenerate.  Any estimate of the mass fraction in
substructure that uses a  simple lens model \markcite{Dalal2002}e.g. {Dalal} \& {Kochanek} (2002) is
highly suspect.  The observational technique used here largely
avoids this important problem, as well as the problems of intrinsic
variability (on the timescale of the typical time-delay between images),
differential extinction, scattering in the radio and microlensing by
stars in the lens galaxy.

The 4-image gravitational lens Q2237+0305, also known as Huchra's lens
after its discoverer \markcite{1985AJ.....90..691H}({Huchra} {et~al.} 1985), has been a testbed for
studying microlensing for many years.  The QSO at redshift $z=1.69$
is lensed by a single spiral galaxy at $z=0.03$ which makes it especially
advantageous for this.  The microlensing of a QSO was first observed by
\markcite{1989AJ.....98.1989I}{Irwin} {et~al.} (1989) in this system and since then it has been
extensively monitored in the optical \markcite{2000ApJ...529...88W}({Wo{\' z}niak} {et~al.} 2000).
Q2237+0305 has also been instrumental in studying the 
structure of the QSO itself.  Constraints on both the size of the
optical emission region \markcite{2001ApJ...548L.127Y,2000MNRAS.315...62W}({Yonehara} 2001; {Wyithe} {et~al.} 2000b)
and the mid-infrared emission region \markcite{2002MNRAS.331.1041W}({Wyithe}, {Agol}, \&  {Fluke} 2002) have been 
established through microlensing. 
In this paper critical measurements of the \hb broad line and \oiii
narrow line ratios are added to the existing information on this lens
which allows us to draw conclusions about the size of the broad line
emission region (BLR) and the size of structures in the lens.

In what follows, if it is not otherwise explicitly stated, the
cosmological model is taken to be a flat one with a cosmological
constant of $\Omega_\Lambda=0.7$ and a Hubble parameter of $H_o=70
\kms\mpc^{-1}$.  The lensing results are often expressed in term of the
critical surface density defined as $\Sigma_{\rm crit}^{-1}(z_s,z_l)=4\pi
GD_{l}D_{ls}/D_s$ where $D_s$ is the angular size distance to the
source, $D_l$ is the distance to the lens and $D_{ls}$ the distance
between the lens and the source.  For the case of Q2237+0305
$\Sigma_{\rm crit}=9.8\times 10^3\msun\pc^{-2}$ in the above cosmology.
The surface density in units of the critical density is denoted $\kappa$.

In the next Section the general method of spectroscopic gravitational
lensing and its special advantages are outlined.  In
Section~\ref{sec:data-reduction} the data reduction is described.  The
measurements of line strengths and magnification ratios are described in
Section~\ref{sec:data-analysis} along with their comparison with
previously published measurements in other wavelengths.  These
measurements are interpreted in Section~\ref{sec:interpretation} in
terms of the structures that could be responsible for them.  A summary
and discussion are given in Section~\ref{sec:discussion}.

\section{The Method of Spectroscopic Compound Lensing}
\label{sec:meth-spectr-comp}

To remove much of the lens model ambiguities, a different method for
detecting substructure lensing was proposed by 
\markcite{MM02}{Moustakas} \& {Metcalf} (2003) which avoids the problems arising from degeneracies in
lens models, differential reddening in the optical images and scattering
in the radio.  This method 
utilizes the prediction that different emission regions of the QSO 
differ in size and thus should be magnified to differing degrees if
substructure on a small scale is present.  Comparing the magnification
ratios in different wavelengths provides information on the structure of
the QSO and the small scale structure of the lens that is independent of
the large scale structure of the lens galaxy and halo.  To describe this method we
will first briefly review what is known about the anatomy of a QSO.

\subsection{sketch of QSO anatomy} 
\label{sec:sketch-qso-anatomy}

The variability timescales of different spectroscopic
features, and the time delays between them, have been used to establish the
size and structure of different emission regions within the unified
model for active galactic nuclei (AGN) and QSOs
\markcite{1993ARA&A..31..473A}(c.f. {Antonucci} 1993). 
The basic picture consists 
of a supermassive black hole (M$_{bh}>10^{6}$M$_{\odot}$) in the core,
surrounded by an accretion disk, which produces the nearly
flat-spectrum continuum light.  The continuum flux varies on very
short timescales, less than a day, and so must be very compact (around
100\,AU or $\sim5\times10^{-4}$\,pc).  It is also known to be
microlensed in the case of Q2237+0305 which puts an upper limit of
2,000\,AU on its size 
\markcite{2001ApJ...548L.127Y,2000MNRAS.318..762W,1990ApJ...358L..33W}({Yonehara} 2001; {Wyithe}, {Webster}, \&  {Turner} 2000a; {Wambsganss}, {Schneider}, \&  {Paczynski} 1990).
Beyond that, there are permitted lines such as the Balmer-series lines
that are very broad ($v_{\rm FWHM}>10^3$\,km\,s$^{-1}$) due to
gravitationally-induced motions.  The characteristic size of this
broad line region (BLR) apparently scales with the intrinsic
luminosity of the host QSO (and therefore with the mass of the central
black hole; \markcite{2000ApJ...533..631K,1999ApJ...527..649W}{Kaspi} {et~al.} 2000; {Wandel} 1999), such
that in luminous QSOs (as opposed to, e.g., Seyfert~1's), the size of
the BLR is on the 
order of six light-months (or $\sim0.1$\,pc).  According to long-term
studies of QSOs, the typical (rest-frame) flux variations in the QSO
continuum are on the order of 10--70\% (though occasional fluctuations
as high as $\sim50\times$ are possible; \markcite{1997ARA&A..35..445U}{Ulrich}, {Maraschi}, \&  {Urry} 1997), while
the BLR variations are smaller by a factor of 2--4 \markcite{1994ApJ...421...34M}({Maoz} {et~al.} 1994).

The narrow emission lines
(e.g. \oiiilam) do not vary in flux significantly over the time span
of several years \markcite{1996ApJ...470..336K}({Kaspi} {et~al.} 1996).  For this reason, they have been
used for calibration in long-term spectroscopic monitoring
observations of low redshift QSOs for reverberation mapping
experiments, by which the time delays between features of different
ionization states may in principle be used to map all the components
that make up QSOs \markcite{1993PASP..105..247P}({Peterson} 1993).  Photo-ionization arguments and
cases where the narrow line region (NLR) is imaged directly, indicate that
the size of the NLR extends out to 100-1000\,pc and is approximately
proportional to the square root of the luminosity in narrow lines.

A QSOs thermal emission in the mid-infrared is believed to come from a
dusty torus that extents from a few parsecs to tens of
parsecs.  This belief largely comes from modeling the radiation
transfer \markcite{1999MNRAS.306..161A}(for example {Andreani}, {Franceschini}, \&  {Granato} 1999) although recently a
more direct lower limit on the size of this emission region has
been made by comparing the microlensing of the optical continuum emission and the
mid-IR \markcite{2002MNRAS.331.1041W}({Wyithe} {et~al.} 2002).

A quasar can emit in the radio on a large range of size scales.  The
central region of a radio loud QSO is usually resolved by VLBI (very long
baseline interferometry) into a core and jet, or a core and lobes on a
scale of $\sim 10\pc$.  The size of the core is limited from below by
the Compton surface brightness limit which is generally in the 10~$\mu$ac range. 

\subsection{the differential magnification ratio (DMR)}
\label{sec:diff-magn-ratio}

Simulations show that if a few percent of the lens surface density is
contained in substructures of mass $\sim 10^7\msun$ then deviations
of $\sim 0.1\mag$ between the magnifications of the NLR and the BLR will
be commonplace \markcite{MM02}({Moustakas} \& {Metcalf} 2003).  In addition, for the case of Q2237+0305
the microlensing scale or Einstein radius for ordinary stars is
particularly large so if 
the BLR is small it could be microlensed.  Our goal here has been to
measure the NLR and BLR
ratios in lens Q2237+0305 and determine if substructure is evident.

To quantify our results we define the differential magnification ratio (DMR)
\begin{equation}
\dmr^{AB} \equiv
\frac{\mu_{NL}^A\mu_{BL}^B}{\mu_{BL}^A\mu_{NL}^B}=\frac{f_{NL}^Af_{BL}^B}{f_{BL}^Af_{NL}^B}.
\end{equation}
where superscripts denote the image and subscripts the QSO emission
region.  The magnification is $\mu$ and the observed flux is $f$.  Note
that the DMR is independent of the intrinsic QSO luminosity, the
intrinsic NLR/BLR ratio and if the NL  
and the BL used are close enough in wavelength it will be essentially
independent of extinction.  In addition, we will define the quantity
$\Delta_{NL/BL} \equiv - 2.5\log(\mu_{BLR}/\mu_{NLR})$ so 
that $\Delta_{NL/BL}^A-\Delta_{NL/BL}^B$ is a measure of the DMR in
magnitudes.  If there is no substructure all three independent DMRs
in a 4--image lens should be close to one 
(a small deviation from one is possible because of the host galaxy's
influence on the NLR; see Section~\ref{sec:nlr-size-effects}).  More
generally the DMR can be used to compare any two measurements of the
magnification ratios to see how well they agree. 

\section{Data Reduction}
\label{sec:data-reduction}
The $J$-band ($\lambda_c \sim 1.25\,\mu$m) spectroscopy of Q2237+0305 was performed with the Cambridge
Infrared Panoramic Survey Spectrograph \markcite{parry2000}CIRPASS Parry {et~al.} (2000) at
the f/16 focus on the 8-m Gemini-South telescope, at Cerro Pachon in
Chile. CIRPASS is a near-infrared fiber-fed spectrograph, connected to a
490 element integral field unit (IFU). The variable lenslet scale was
set to $0\farcs 36$ diameter, and the hexagonal lenslets are arranged in
the IFU to survey an approximately rectangular area of $13\farcs0\times
4\farcs7$. The detectors is a 1k$\times$1k Hawaii-I HgCdTe Rockwell
array. CIRPASS can operate in the range $0.9 - 1.8\,\mu$m, and a 400l/mm
grating was used which produced a dispersion of 2.25\,\AA /pix. The
grating was tilted to place the wavelength range $\lambda\lambda
11300-13600$\,\AA\ on the detector, covering most of the $J$-band
transmission window out to the atmospheric absorption between the $J$-
and $H$-bands. A filter at $1.67\,\mu$m blocked out redder wavelengths to
reduce the dominant near-infrared background contribution. The detector
pixels do not quite critically sample the spectral resolution
(unresolved sky lines have FWHM=1.7 pix), and the resolving power is
$\lambda\,/\,\Delta\lambda_{\rm FWHM}\,=\,3500$.

The observations were made on the night of UT 2002 August 10 during the
Director's discretionary time instrument commissioning/IFU-demonstration
science on Gemini-South. The observations spanned the airmass range
$1.2-1.8$. A total of 3\,hours integration was obtained, split into
individual integrations of 30\,min per pointing, although the array was
read non-destructively every 10\,min (i.e., 4 times per pointing,
including an initial read of the bias level). Each of these `loops'
comprised 10 multiple reads of the detector, which were averaged to
reduce the readout noise (30$e$- per read) by $\sqrt{10}$. These averages
of each loop were subtracted from the average of the next loop of
non-destructive reads, to form 3 sub-integrations of 10\,min at the same
telescope pointing. Comparison of these sub-integrations enabled cosmic
ray strikes to be flagged, and a combined frame of 30\,min was produced
by summing the three sub-integrations, while masking pixels affected by
cosmic rays.  The seeing was $\approx 1\farcs0$ FWHM, and the long axis
of the IFU was set to 130 deg East of North. Windshake affected the
guiding, and 2 loops were lost through severe trailing of the
images. These were removed from the subsequent data reduction, so the
final science integration was 160\,min. In total, six overlapping
pointings were surveyed, each covering all of the Einstein cross:
positions were not repeated to improve background subtraction and reduce
the impact of bad pixels.  We `nodded' the target from the top {\em (A)}
to the bottom {\em (B)} of the IFU every 1800\,s, and introduced a small
sub-dither of 2 lenslets between integrations.  An initial subtraction
of sky and dark current was performed using this beam-switching: for
each exposure at location {\em A}, the average of the three spectra
taken at {\em B} was subtracted, and vice versa. There was no overlap of
the extended emission from the Einstein cross between positions {\em A}
and {\em B} (a chop of $\approx 8''$). Known bad pixels were also
interpolated over at this stage.

The 490 fibers span the 1k detector, with 2 pixels per fiber. A lamp was
used to illuminate 10 calibration fibers immediately before the
observations, in order to accurately determined the position of the
fibers on the array and to focus the spectrograph at the desired
wavelength range. There is significant cross-talk between adjacent
fibers and we use an optimal-extraction routine\footnote{The CIRPASS
data reduction software is available from
http://www.ast.cam.ac.uk/\~{}optics/cirpass/docs.html}
\markcite{Johnson2003}(Johnson {et~al.} 2003) to determine the spectrum of each fiber, for each
pixel on the array solving the contribution of flux from adjacent fibers
(knowing their relative positions).

Immediately after the science integrations, spectral flat fields were obtained
(exposures of the illuminated dome), and these were also
optimally-extracted and their average flux normalized to unity.
The extracted science data was then
flat-fielded through division by the extracted normalized dome lamp
spectra, calibrating the response of each individual fiber.

Wavelength calibration was achieved with an Argon arc lamp observed
following the science exposures. Again, the individual fiber spectra
were optimally extracted, and for each fiber a cubic fit was performed
to the centroids of 40 arc lines, leaving {\em rms} residuals of
0.2\,\AA . The 490 fiber spectra stacked on the slit were rectified
(mapped to the same wavelength) with a 5th order polynomial
transformation in $x$ and $y$. This transformation was applied to the
beam-switched, extracted, flat-fielded science frames. A higher-order
background subtraction was then applied to this rectified data, to remove
sky residuals from the beam switching caused by variation in the OH-line
intensity.

The rectified, background-subtracted 2D spectra (with the dimension of
wavelength and fiber number along the slit) were then converted to 3D
data cubes, arranging the fibers into their physical positions in 2D on
the sky, with wavelength being the third dimension. Because of the hexagonal
close packing arrangement of the lenslets, alternate rows are offset by
half a lens: in making the cubes this was accounted for by regridding
each lenslet by two in the spatial dimensions (i.e. each lenslet is 2x2
sub-pixels of diameter $0\farcs 18$).  These six data cubes (from the
six pointings of the telescope) were registered, shifting them spatially
to the same location on the sky, and combined with a percentile clipping
algorithm, rejecting bad pixels and residuals from cosmic ray strikes
discrepant by more than $3\,\sigma$.

Flux calibration was obtained through observations of the standard star
HIP036096 ($V=8.89$\,mag, type B2V), taken on the same night at similar
airmass. The standard star spectrum was reduced in the same way as the
science data, and the total flux found by summing $4\times 4$ lenslets
(1.5\,arcsec diameter). Normalizing through division by this spectrum 
corrected for the severe atmospheric absorption at $>1.33\,\mu$m. 
The total effective throughput was determined to be $\sim 8$ per cent on
the sky for the middle of the $J$-band.

\begin{figure}[t]
\centering\epsfig{figure=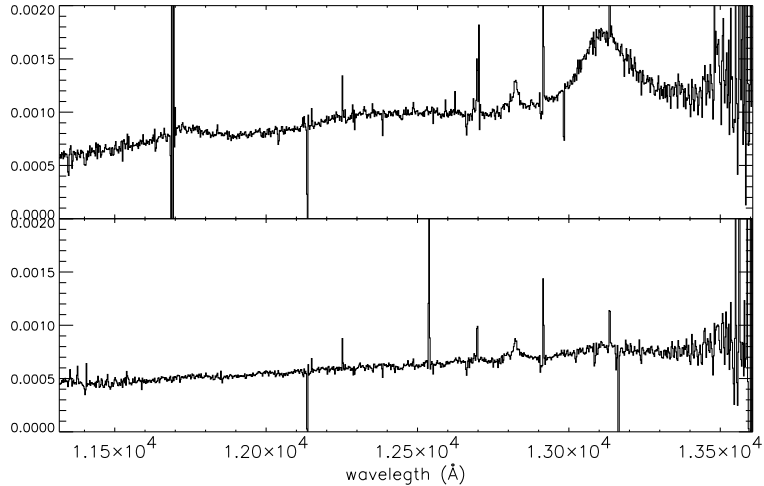,height=3.0in}
\caption[the]{\footnotesize The full CIRPASS spectrum at the center of
  QSO image~A (above) and at the center of the lens galaxy (below).  The
  broad H$\beta$ line at 13097~\AA\, is clear in the QSO spectrum and
  in the galaxy spectrum there is only a hint of it.  It is also clear
  that the continuum is steeper in the QSO and the \feii ``bulge'' is
  apparent in the QSO spectrum near 12300\,\AA.  The prominent line at 1.28~\micron is Pa$\beta$ from the calibration star.}
\label{fig:full_spec}
\end{figure}

\section{Data Analysis}
\label{sec:data-analysis}

\subsection{line strengths}
\begin{figure}[t]
\centering\epsfig{figure=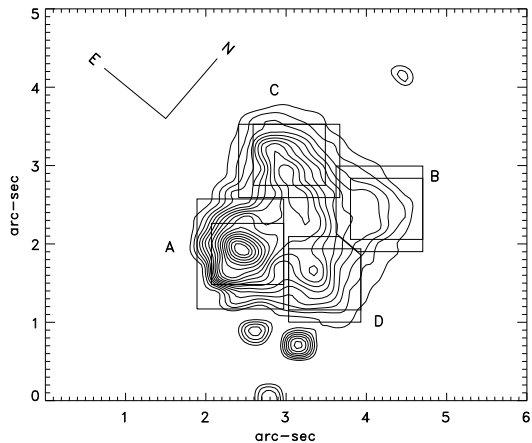,height=2.5in}
\caption[]{\footnotesize Map of the integrated flux between
  11305.1\,\AA\, and 13613.7\,\AA (or 4203\,\AA to 5061\,\AA in the
  QSO's rest-frame).  The image is in the CIRPASS spixel
  coordinates which are rotated clockwise $50^\circ$ from $\delta$--R.A.
  coordinates.  The squares mark the apertures used to measure the QSO 
  fluxes as discussed in \S~\ref{sec:meas-magn-ratios}.  The
  conventional image labels are shown next to each aperture.
  Note that image~B (the right most) is not prominent above the lens
  galaxy.  The pixels are 0.18\arcsec\, in the 
  horizontal direction and 0.156\arcsec\, in the vertical (the difference
  is due to the hexagonal packing of the CIRPASS lenslets).  The two
  spots below the central image are caused by cosmic rays.
}
\label{fig:total_flux}
\end{figure}

\begin{figure}[t]
\centering\epsfig{figure=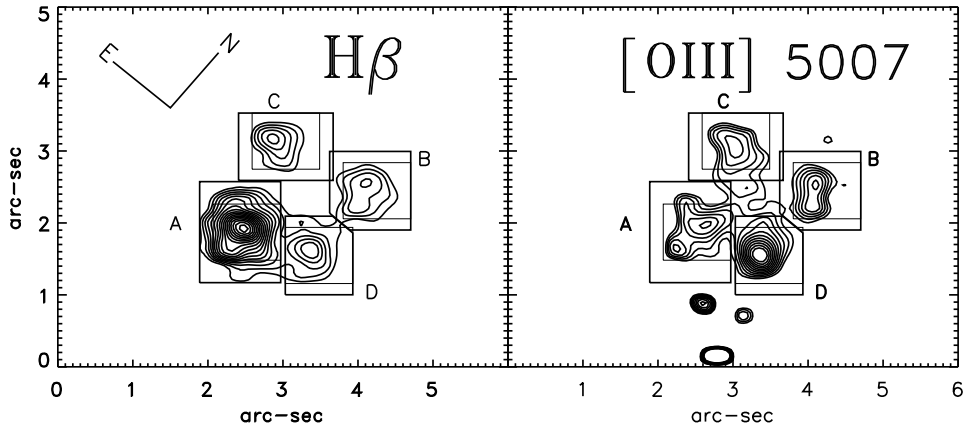,height=2.75in}
\caption[]{\footnotesize A map of H$\beta$ and \oiii 5007 at the
  redshift of the QSO with the continuum subtracted spixel-by-spixel.
  The four QSO images are
  more distinct than in the total flux,
  Figure~\ref{fig:total_flux}. Image~B is clearly apparent here and the 
  lens galaxy is not.}
\label{fig:Hb_OIII}
\end{figure}

To extract the line fluxes we fit a local continuum spectrum plus lines to
each ``spixel'' in the image (i.e. the spectrum from each fiber
corresponding to a unique spatial position).  
Example spectra from two spixels are shown in
Figure~\ref{fig:full_spec}.  We are interested in the relative integrated flux
within a line which should not depend critically on modeling the line
shapes in detail.  (In fact it is possible that microlensing might
affect the line profile in detail without greatly affecting the total
flux in the line in which case a fit to a smoother line profile would
be preferable.  See Section~\ref{sec:hbeta-line-profiles}.)  We start by
fitting a local continuum that is linear in wavelength and a simple first
order Edgeworth expansion\footnote{The Edgeworth expansion is a generalization
of the Gaussian distribution where each addition term effects 
a higher moment of the distribution.  The first order expansion has a
skew.} to the H$\beta$ line in the nine spixels around image~A, the
brightest image.  This fit is done only in the range 12658--13424~\AA\,
(4706--4990~\AA\, in QSO rest frame) to avoid the \feii complex as much
as possible.  The shape of the H$\beta$ line in image~A is then used as
a template for all the other spixels. The spectrum at each spixel in the
array is fit to a continuum level and slope along with a line strength.
The result is displayed in Figure~\ref{fig:Hb_OIII}.  In fitting the
line and continuum several sky lines are masked out and along with the
calibration star's Pa$\beta$ at 12820\AA, see
Figure~\ref{fig:full_spec}.  The shape and level of the \feii complex
varies greatly between QSOs and is thus very difficult to subtract
\markcite{1992ApJS...80..109B}(see {Boroson} \& {Green} 1992).  In all cases however \feii
contributes very little to the flux between 12658~\AA\, and the H$\beta$
line.  There are three broad \feii lines blueward of H$\beta$ at
$\lambda\lambda4924,5018,5169$~\AA\, which are not identified in our spectra.
The 4924~\AA~ line could conceivably cause us to overestimate the slope
of the continuum and underestimate the strength of H$\beta$.  We can
estimate the maximum effect this could have by noting that the height
of this line is always below the height of the $\lambda\lambda4450-4700$
\feii bulge which we can measure.  This contribution to the H$\beta$
strength is rather small.  We include it in the H$\beta$ flux errors
instead of attempting to subtract it.

The full--width--half--maxima (FWHM) of the H$\beta$ lines are 4100,
3500, 3600 and 3600$\pm 100\kms$ for images~A, B, C and D respectively -
all consistent with each other except image~A.  
The skewnesses of the images are all small and consistent with each other.
There is further discussion of the change in the shape of the  H$\beta$
between images in Section~\ref{sec:hbeta-line-profiles}.

The sky absorption near the 5007\AA~ and 4959\AA~ \oiii lines is large
and sky line subtraction makes any fitting difficult.  The 4959\AA~ line is
not detected anywhere in the field possibly because of confusion with
the \feii\,$\lambda4924$ line.  The 5007\AA~ line is very close to a
subtracted sky line to its redward side, and to its blueward the noise
introduced by flux calibration and high sky absorption is too large
to accurately fit a continuum.  To avoid these difficulties we fit the
continuum in the range 13278--13469\AA~(independently of the continuum
used in the \hb fit) then subtract this from the sum of the flux in 7 pixels 
(in wavelength) on top of the line (rest frame 5006-5010\AA).  This
avoids the sky line.  No fitting of the line shape is done.

The maps of line strengths in Figure~\ref{fig:Hb_OIII} show the four QSO
images distinctly and residual flux from the lens galaxy is at a low
level although there are bridges between 3 of the images in \oiii at a level
of 25\% of the maximum.  These connections go through the center of the
lens galaxy and no arc connecting images is detected.
The \oiii NLR is consistent with being point-like at our resolution, $\sim
0.6\arcsec$ in contrast to the C\,{\scriptsize III}]\,$\lambda1909$ NLR
observed by \markcite{1998ApJ...503L..27M}{Mediavilla} {et~al.} (1998) which shows an arc connecting
images A, D and B.  

To evaluate the statistical errors in our line strengths we take the
part of the IFU field that is well away from the lens system where we
expect no line emission and go through the same continuum and line
fitting procedure.  We find that the average line strengths in 1036 such 
spixels is consistent with zero showing that there is no significant
bias in our line fitting algorithm.  The variance in the line strengths 
among these spixels is used as the statistical noise.  There should be no
significant signal-dependent noise since the data is dark current limited.

\subsection{measuring magnification ratios}
\label{sec:meas-magn-ratios}

To find the total flux within a line coming from an image we must choose
an aperture around each image.  We use two choices for apertures.  First
we use square boxes five half-lenses wide
($0.90\arcsec\times0.78\arcsec$) centered on the QSO image positions.
This is close to the expected size of the PSF.  The peak in the emission
is found to be the center of image~A and the other image positions are
set using the high precision HST relative image positions \markcite{CASTLES}(Kochanek {et~al.} 2000).  These are
shown in Figures~\ref{fig:total_flux} and \ref{fig:Hb_OIII}.  To make certain
that we are capturing all the flux from each image we also extend the
apertures as shown in the figures.  The differential magnifications are
given in Table~\ref{table} along with the separate NLR and BLR flux
ratios.  The errors given are the statistical errors discussed above.
There is little change to the DMRs when the
apertures are extended further without overlapping each other signifying that
leakage of flux outside of the apertures is not significant.

Because the DMRs depend on
which image is chosen to normalize them the important features to look
for are the spread in the DMRs, how consistent they are with each other,
and whether there is one outlier indicating that one particular image is
being effected.  The normalizing image's DMR -- $\Delta=0$ -- should be
included in this comparison.  In table~\ref{table} image~A is used to
normalize the DMRs.  For the other images the DMRs are negative and inconsistent with
zero, but they are roughly consistent with each other.  This indicates
that the BLR in image~A is anomalously bright.
The simple magnification ratios in the BLR and those in NLR are also consistent
with each other except those including image~A which indicates that the
differential extinction is not large.
\begin{table}[t]
\begin{center}
\begin{tabular}{llccc}
& Image  & B/A & C/A & D/A \\
\hline
small apertures & $\Delta_{\rm NL/BL}$   & $-0.82\pm0.06$  &
  $-0.89\pm0.06$ & $-0.93\pm0.05$ \\
 & $\Delta_{\rm NL}$   & $-0.23\pm0.06$ & $-0.14\pm0.06$ &
$0.09\pm0.05$   \\
 & $\Delta_{\rm BL}$  & $-1.06\pm0.02$  & $-1.03\pm0.02$ &
$-0.84\pm0.01$ \\ 
extended apertures 
& $\Delta_{\rm NL/BL}$ & $-0.74\pm0.05$  & $-0.87\pm0.05$ &
$-0.81\pm0.05$ \\
 & $\Delta_{\rm NL}$  & $-0.31\pm0.05$ & $-0.23\pm0.04$ &
$-0.21\pm0.04$  \\ 
 & $\Delta_{\rm BL}$  & $-1.05\pm0.02$ & $-1.11\pm0.02$ &
$-1.02\pm0.01$  \\ 
\end{tabular}
\caption{\footnotesize  {\bf Differential Magnification Ratios and
  Magnification ratios in magnitudes:} These
  are normalized so that $\Delta_{\rm NL/BL}^A=0$.  Negative values
  indicate that image~A is brighter in the simple ratios and that the
  BLR of image~A is brighter in the DMRs. \label{table}}
\end{center}
\end{table}

\subsection{Continuum}

It would be useful to compare the BLR and NLR fluxes with the continuum
in our data. Extracting the QSO continuum flux requires subtracting the lens galaxy
contribution and all the contributions from lines.  An attempt was made
to model the HST NICMOS image of the lens 
galaxy and degrade it to the PSF of the Gemini/CIRPASS data.  We
concluded that because of wind-shake, the PSF of our data was not well
enough characterized to accomplish this to the accuracy we would like
for this paper.

\section{comparison with radio and mid-infrared ratios}
\label{sec:comp-with-radio}

\begin{figure}[t]
\centering\epsfig{figure=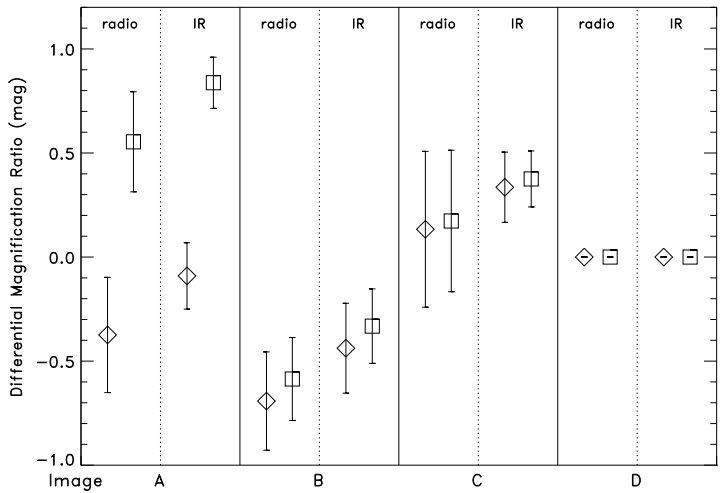,height=2.75in}
\caption[]{\footnotesize Here are the DMR with respect to the radio
  and mid-IR fluxes normalized to image~D.  The NLR DMRs
  (i.e. $\Delta^i_{\rm NL/radio}-\Delta^D_{\rm NL/radio}$ and
  $\Delta^i_{\rm NL/IR}-\Delta^D_{\rm NL/IR}$) are marked by 
  diamonds and the BLR DMRs by squares.  The radio is at 3.6~cm and the
  mid-IR is a combination of 8.9~\micron and 11.7~\micron observations.
  The published 1-$\sigma$ error bars are shown.}
\label{fig:DMR_radio_IR}
\end{figure}
Two of the goals of this work are to eliminate the uncertainty inherent
in radio flux ratios due to scattering by ionized gas and to eliminate
differential extinction uncertainties inherent in optical or near infrared
observations.  However, it will be shown that in this case these effects
are small.  To do this and gain more information about substructure in
the lens we construct DMRs by combining other observations with the NLR
and BLR measurements.  
Since the sizes of the radio and mid-infrared emission regions are believed to
be intermediate to those of the BLR and NLR this can give us additional
information about the structure causing the anomalous DMR at the expense
of these additional uncertainties.

\begin{figure}[t]
\centering\epsfig{figure=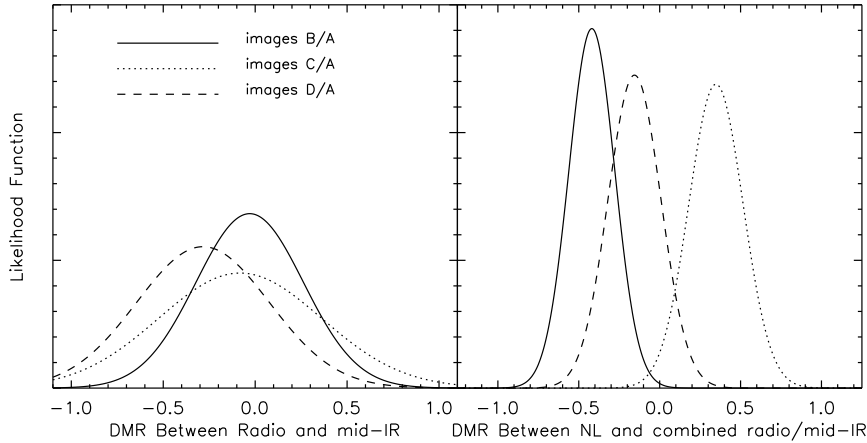,height=3.5in}
\caption[]{\footnotesize The likelihood functions for differential
  magnification ratios.  In the left panel is the DMR between the mid-IR
  and radio ratios with respect to image~A (DMR=0 means the ratio of
  mid-IR to radio is the same as in image~A) using the published errors.
  All three DMRs are compatible with zero.  In the right panel it is
  assumed that the magnification ratios in mid-IR and radio are the same
  and they combined into one set of measurements.  The likelihood
  function between this measurement and the NLR measurements are
  plotted.  The NLR errors are twice those reported in Table~\ref{table}
  to be conservative.  The DMRs of image~C and D are consistent with
  zero, but image~B has a low DMR meaning that the radio/mid-IR is
  anomalously bright with respect to the NLR in this image. 
}
\label{fig:dmr_comparison}
\end{figure}

The radio fluxes in Q2237+0305 have been measured using VLA (Very Large
Array) by \markcite{1996AJ....112..897F}{Falco} {et~al.} (1996) at 3.6~cm, and
\markcite{2000ApJ...545..657A}{Agol}, {Jones}, \&  {Blaes} (2000) have observed it in the mid-infrared,
8.9~$\mu$m and 11.7~$\mu$m.  Figure~\ref{fig:DMR_radio_IR} shows the
DMRs with respect to the radio and mid-IR measurements for both the NLR
and the BLR.  The uncertainties in these measurements are considerably
larger than those in our line measurements, but important conclusions
can be drawn from them.  The BLR in image~A is 
anomalously bright with respect to the other measurements while the NLR in
image~A is consistent with being magnified to the same extent as both
the radio and mid-IR.  The consistence in the mid-IR and radio ratios is
further demonstrated in the left panel of
Figure~\ref{fig:dmr_comparison}.  This suggests that the radio and the
mid-IR emission regions are being lensed by the same structure and thus
are similar in size which is in agreement with theoretical expectations,
but, to our knowledge, has never been measured directly. It is also seen
in Figure~\ref{fig:DMR_radio_IR} and more clearly in the right
panel of Figure~\ref{fig:dmr_comparison} that both the radio and the
mid-IR do not agree with the NLR -- the image~B NLR is relatively dim
and the image~C NLR is relatively bright.  The difference between the largest
DMR and the smallest DMR, the spread, is $0.77\mag$.  There is no clear
outlier in the DMRs suggesting that the spread is not the result of an
anomaly in a single image.
Table~\ref{big_table} lists the DMRs comparing the combined radio/mid-IR
with the NLR and BLR data.  

\begin{table}[t]
\begin{center}
\begin{tabular}{lcccccc}\footnotesize
& \multicolumn{3}{c}{\bf narrow line region} & \multicolumn{3}{c}{\bf
    broad line region}  \\
image & B/A & C/A & D/A & B/A & C/A & D/A \\
\hline
mid-IR/radio &  $-0.42\pm0.11$ & $0.35\pm0.15$ & $-0.16\pm0.15$ 
             &  $-1.16\pm0.10$ & $-0.53\pm0.15$ & $-0.97\pm0.14$\\
\end{tabular}
\caption{\footnotesize  
The differential magnification ratios comparing the combined radio and
  mid-IR magnification ratio measurements to the NLR and BLR measurements.  This is
  $\Delta_{\mbox{radio,mid-IR}}-\Delta_{\rm NL}$ and
  $\Delta_{\mbox{radio,mid-IR}}-\Delta_{\rm BL}$.  The NLR and BLR
  errors are not doubled here.  The NLR B to A and C to A DMRs disagree
  by 0.77~mag.
\label{big_table}}
\end{center}
\end{table}

The differential colors for the images of Q2237+0305 are in order
$\Delta E(B-V)= -0.10\pm 0.03$, $-0.17\pm0.03$, $0.01\pm0.03$ and $0$ 
\markcite{1999ApJ...523..617F}({Falco} {et~al.} 1999).  This implies that extinction would make
the NLR in image~B relatively dimmer than the others which is the
opposite of what is found.  Also, the expected magnitude of the
extinction is relatively 
small in comparison to the DMRs observed -- $A(1.3\mu m)/E(B-V) =
0.75\mag$ for the Milky Way giving a change in the DMR between images~A
and B of $0.05\mag$.  This does not exclude the possibility that
the extinction is very patchy so that the continuum extinction is not
appropriate for the NLR, but the small differences between the colors of
the images argue against this being a large effect.

The simplest interpretation of the anomalous DMRs in
Figure~\ref{fig:dmr_comparison} is that the mid-IR and radio emission
regions are larger than the BLR and whatever is causing the extra
magnification of the BLR in image~A is too small to significantly change
the magnification of the mid-IR, radio or NLR.  Then there is some
larger structures that are causing the NLR ratios to be incompatible with the
mid-IR and radio ratios.

\section{Interpretation}
\label{sec:interpretation}

There are three possible factors contributing to the anomalous differential
magnification ratios we observe.   Microlensing of
the BLR by stars in the lens galaxy could be changing its magnification
ratios relative to those of the NLR, radio and mid-IR.  This is
discussed in Section~\ref{sec:microlensing}.  The size of the
NLR could be big enough with respect to the host lens that its 
magnification differs from that of the other emission regions even
without what would normally be classified as substructure.  This is ruled out in
Section~\ref{sec:nlr-size-effects}. 
Finally, larger scale substructure (possibly made
of CDM) either in the lens galaxy or along the line of 
sight is causing the mismatch between the NLR and radio/mid-IR
magnification ratios.  This is termed millilensing and is investigated
in Section~\ref{sec:cold-dark-matter}.  

An additional consideration that can be immediately ruled out is that
the time delay between images is longer than the typical timescale for
variations in the source.  Any reasonable lens model for Q2237+0305
gives time delays that are very short, typically less than a day.  Recently,
a tentative measurement of the time delay has been made using X-ray
observations yielding $2.7^{+0.5}_{-0.9}$~hours between image~A and B
\markcite{2003ApJ...589..100D}({Dai} {et~al.} 2003).  This is well below the timescale on which
the QSO is expected to vary in any of the wave--bands discussed in this paper.

\subsection{Microlensing}
\label{sec:microlensing}

Q2237+0305 is exceptional among QSO lens systems in that its lens
galaxy is very close to us (z=0.03).  This makes the BLR particular
susceptible to microlensing for two reasons.  First, the Einstein radius of the
host galaxy in physical units is roughly $\propto D_{ls}D_{l}/D_s$ which
reaches its maximum close to half way to the source.  For this reason
the images form closer to the center of the galaxy in Q2237+0305 where
the star density is particularly high.  In addition, the size of a
star's Einstein radius relative to the size of the source goes as
$\sqrt{D_{ls}D_{s}/D_l}$ and so is exceptionally large for Q2237+0305.
There is ample evidence that the QSO's optical continuum is being
microlensing in image~B of Q2237+0305
\markcite{1989AJ.....98.1989I,1996A&A...309...59O,1991AJ....102...34C,2000ApJ...529...88W,2000MNRAS.315...62W}({Irwin} {et~al.} 1989; {Ostensen} {et~al.} 1996; {Corrigan} {et~al.} 1991; {Wo{\' z}niak} {et~al.} 2000; {Wyithe} {et~al.} 2000b).  Monitoring has shown
that in the optical, image~B went from being the second brightest image
(hence the name) to being tied for the dimmest image over a two year
period.  We find image~B to be dimmest in the NLR and the BLR.
This gives some hint that the BLR magnification is correlated with the
microlensing of the optical continuum.

Further evidence of microlensing of the BLR is in
Figure~\ref{fig:DMR_radio_IR} where our data is compared to other
data.  The mid-infrared emission region is believed to be not much
larger than the BLR.  The lack of strong anomalies in the mid-IR/radio
DMRs suggests that the substructure is very small.  In addition, the
large anomaly in the BLR/NLR image~A DMR implies compact and numerous
substructures are responsible.  Microlensing by stars seems like the
most straightforward explanation.

The microlensing interpretation is also consistent with theoretical
expectations for this system.  An upper limit on the standard deviation of the
magnification ratio caused by microlensing for a finite sized source is
$\delta m \simlt 3 \frac{\sqrt{\kappa_*}}{|1-\kappa+\kappa_*|} 
\frac{\theta_{\rm E}(M_*)}{\theta_s}$ 
\markcite{1997A&A...325..877R,1991A&A...250...62R}({Refsdal} \& {Stabell} 1997, 1991) where  $\theta_{\rm
  E}(M_*)$ is the Einstein radius for a star of mass $M_* = \langle
M^2\rangle/\langle M\rangle$ , the averaging is done over the
distribution of star masses, and $\kappa_*$ is the dimensionless surface
density in stars.  The surface density in smoothly distributed dark
matter is $\kappa-\kappa_*$.  In this case $\theta_{\rm E}(M_*) = 0.089 \pc/D_s
\sqrt{M_*/\msun}= 3.63\times 10^{-6} \sqrt{M_*/\msun}\mbox{ arcsec}$.
\markcite{2002MNRAS.334..621T}{Trott} \& {Webster} (2002) have modeled the mass distribution of
Q\,2237+0305 using the QSO image positions and a variety of direct
observations of the lens galaxy.  Their best fit model gives
$\kappa-\kappa_*\simeq 0.05$ and $\kappa_*\simeq 0.04$ at the radius of the
images.  For a realistic value of $M_*=0.5 \msun$ this gives $\delta m \simlt
0.04\pc/R_{\rm NLR}$ where $R_{\rm NLR}$ is the physical size of the BLR.

The size of the BLR in some QSOs and Seyfert galaxies can be estimated
using reverberation mapping.  Combining the radius--luminosity relation
and broad line width--luminosity relation of \markcite{2000ApJ...533..631K}{Kaspi} {et~al.} (2000)
gives $R_{\rm NLR}\simeq 0.03 (4108\kms/v_{\rm FWHM})^{2.58}\pc$ although
there is significant scatter about this relation.  With the
H$\beta$ line widths reported in \S\ref{sec:data-analysis} this gives
$R_{\rm NLR} \sim 0.05\pc$, small enough that the the observed
magnification of image~A can be explained by microlensing. 

The time scale for changes in the microlensing magnification
is set by the size of the BLR and the velocity of the stars in the
galaxy relative to the image.  The BLR is of order a light--month in
size and typical speeds for the galaxy and rotation of the its disk are 
$\sim 200\kms$.  This gives a time scale of $\sim 100\yr$.  This is in
contrast to microlensing of the visible continuum where the time scale
is about 100 times smaller \markcite{2000MNRAS.315...62W,2000MNRAS.318..762W}({Wyithe} {et~al.} 2000b, 2000a).

\subsubsection{H$\beta$ Line Profiles}
\label{sec:hbeta-line-profiles}

\begin{figure}[t]
\centering\epsfig{figure=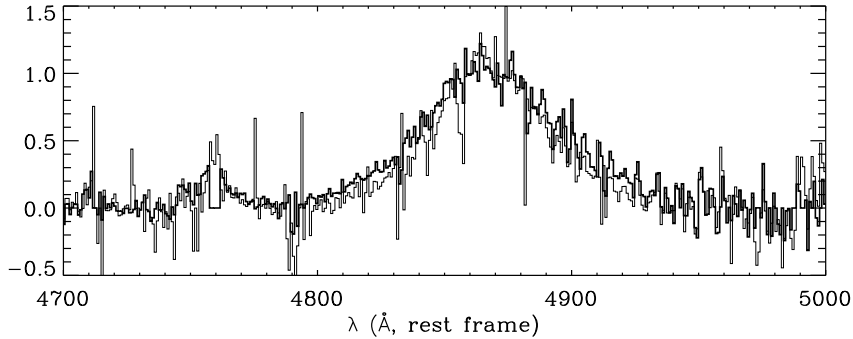,height=3.5in}
\vspace{-1in}
\caption[]{\footnotesize Comparison of the H$\beta$ line in image~A
  (dark curve) to that in image~B (light curve).  The line in image~A is
  slightly wider than the line in image~B.  This could be caused by a
  blending of the emission from the NLR and the BLR combined with
  differing magnifications or by microlensing of kinetically distinct
  regions of the BLR.
}\label{fig:Hbeta_lines}
\end{figure}

Several groups have investigated the possibility of detecting
microlensing and learning something about the structure of the BLR
through distortions in the profile of the broad lines \markcite{1988ApJ...335..593N,1990A&A...237...42S,2002ApJ...576..640A}({Nemiroff} 1988; {Schneider} \& {Wambsganss} 1990; {Abajas} {et~al.} 2002).  
If the BLR has a coherent, large scale velocity structure (such as rotation
or uniform infall) microlensing could change the shape of a broad line
in the image that is being microlensed.  This effect comes about because
the magnification can vary across the image of the source and the
emission region for a narrow range of wavelength can be much smaller
than the whole NLR.  It was generally thought that if
microlensing would have an insignificant effect on the total
magnification of the BLR of a bright QSO. 

We find that the H$\beta$ line is wider in image~A with respect to the
other images.  Figure~\ref{fig:Hbeta_lines} shows the line in image~A
and B with the continuum subtracted and normalized so they have the same
height.  The change is not large and there is no significant evidence
that the asymmetry of the lines are different in the different images.
In appendix~\ref{sec:contr-wavel-depend} an upper limit to the 
contribution a wavelength dependent component to the magnification is
derived.  More than half of the anomalous magnification in image~A must
be caused by a magnification that is uniform over the velocity structure
of the H$\beta$ emission region.  

The extra width of the image~A H$\beta$ line can be interpreted as
either caused by the microlensing of a small scale velocity structure in
the BLR or as a sign that there
is a narrow--line component to the H$\beta$ that is not being magnified
as strongly in image~A as the broad--line component.  The microlensing
explanation would not in general produce a symmetric broadening like the
one we see, but it is possible that it would.  We conclude that there is
no strong evidence for coherent large scale motion in the H$\beta$ BLR,
but that this possibility cannot be ruled out.

\markcite{1989ApJ...338L..49F}{Filippenko} (1989) found that the broad MgII $\lambda$2798 line had a
slightly asymmetric distortion in its profile between the different
QSO images of Q2237+0305 which could be attributed to microlensing.  It is
known that the BLR for high ionization states is significantly smaller
than the H$\beta$ BLR \markcite{1999ApJ...521L..95P}(see {Peterson} \& {Wandel} 1999, for an illustration of
  this) thus more susceptible to microlensing and possibly
more coherent in its velocity structure.

\subsection{NLR size effects}
\label{sec:nlr-size-effects}
\begin{figure}[t]
\centering\epsfig{figure=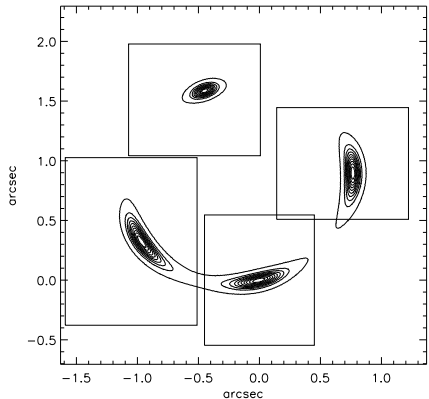,height=2.25in}
\caption[]{\footnotesize A model of Q2237+0305 with the regions used for
  measuring the magnification ratios shown.  In this case the source
  has a Gaussian profile of size $R_{0.1}=526\pc$ (full width for 10\% peak
  surface brightness).  The lowest contour is at 1\% of the peak surface 
  brightness and they go up in intervals of 10\%.  It can be seen that
  differences in the  $\Delta_{NL/BL}$'s from zero are not caused by
  loses outside of the integration regions.
}
\label{fig:finite_map}
\end{figure}
\begin{figure}[t]
\centering\epsfig{figure=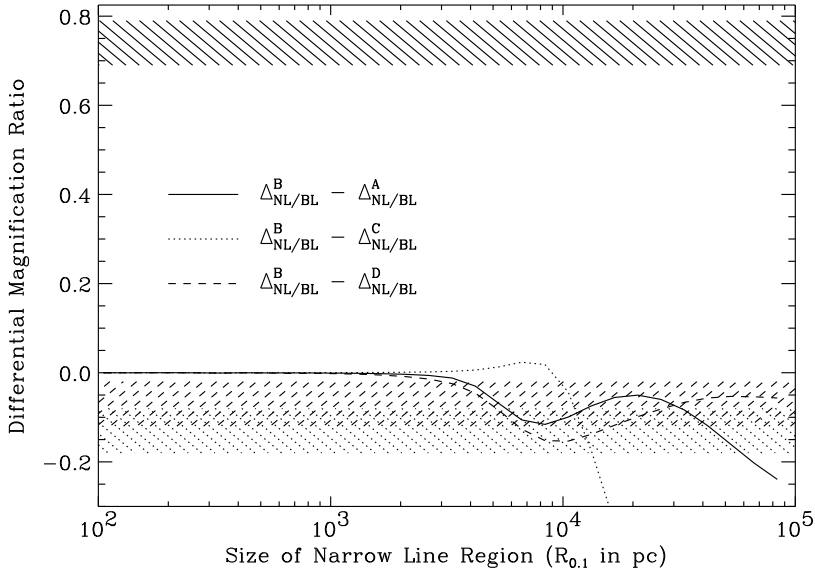,height=3.25in}
\caption[]{\footnotesize The DMRs with respect to image~B are plotted
  as a function of source size, $R_{0.1}$.  These are calculated using
  the lens model and the extended apertures discussed in the text.  The
  source profile is Gaussian in this example.  The hashed regions are
  within $1\sigma$ of the measured values.}
\label{fig:finite_mag}
\end{figure}

There are two ways in which the size of the NLR might change the NL/BL
ratio between images without substructure.  First, if the effective size
of the NLR is close to the size of the aperture used to measure the flux
the NLR/BLR ratio will be affected.  Because the sizes of the NLR are
different in each image it is possible some of the flux could leak out
of the aperture causing the more magnified images to have lower NLR to
BLR ratios.  This possibility can be eliminated by increasing the size
of the apertures and seeing in the NLR fluxes go down as is the case for
our extended apertures.

The second possibility is that the NLR is large enough that the
magnification cannot be considered uniform over the whole image and thus
its total magnification will differ from that of the smaller BLR.  
To address this issue we model the lens and the NLR for Q2237+0305
and measure the ratios in mock data.  The lens model consists of a
elliptical halo with a power-law radial profile, a exponential disk and
an external shear.  The model reproduces the image positions and the
radio magnification ratios.  The same apertures are used in the
simulations and with the data.  Figure~\ref{fig:finite_map} shows an
example of one such simulation.  We use two models for the surface
brightness profile of the NLR -- an exponential and a Gaussian.  The
Gaussian is meant to represent the case of a NLR with a sharp edge while
the exponential is more diffuse.  The profile of the NLR of NGC 4151,
probably the best studied example in a Seyfert galaxy, is roughly
exponential although it does show significant irregularities and
asymmetries \markcite{1994A&A...291..351R}({Robinson} {et~al.} 1994).

The size of the NLR in C\,{\scriptsize III}]\,$\lambda1909$ was 
inferred from modeling the arc connecting images A, B and D by
\markcite{1998ApJ...503L..27M}{Mediavilla} {et~al.} (1998).  They estimate the size of the BLR in this
wavelength to be $340-570 h^{-1}\pc$ where this is the total extent of the
observed emission.  Direct HST measurements of QSOs in \oiii indicate that
the the NLR could be a few kpc in total extent \markcite{2002ApJ...574L.105B}({Bennert} {et~al.} 2002).

Figure~\ref{fig:finite_mag} shows the results of these simulations along
with the observed DMRs.  The size of the source is measured by $R_{0.1}$
which is the radius at which the surface density drops below $10\%$.
For comparison $1\arcsec= 24.5 h_{\rm 65}^{-1}\kpc$ on the source plane and $885
h_{\rm 65}^{-1}\pc$ on the lens plane.  For a Gaussian source profile
the DMR does not differ greatly from one (or zero in magnitudes) if
$R_{0.1}\simlt 4\kpc$.  For a exponential profile the same is true for
$R_{0.1}\simlt 2\kpc$.  And when the NLR is large enough to change
the DMR it is in the opposite direction to the observed A to B DMR.  We
conclude that the finite size of the NLR could account for the low (but
not very low) C to B and D to B DMRs, but not for the high A to B DMR.

\subsection{Millilensing}
\label{sec:cold-dark-matter}

The disagreement between the NLR and radio/mid-IR magnification ratios
is most easily explained by the presents of substructure with a mass $m
\gg \msun$.  We shall see that the quantity of substructure required to
make this discrepancy probable makes substructure made of CDM a more
likely explanation than substructure composed of ordinary gas and stars
such as dwarf galaxies or globular clusters.  To determine the level and
properties of the substructure that are compatible the measured DMRs we
carry out extensive lensing simulations.

\subsubsection{simulations}
\label{sec:simulations}

For the results that follow, the magnification distributions were
calculated using an adaptive grid code.  The strategy is to first calculate
the image positions, through the lens equation, on a coarse grid
over the image plane.  The deflection angle is found by summing over all
the subclumps in this initial region.  The grid is then refined to cover
smaller regions surrounding each of the images.  The grid is made finer
when the image begins to cover a large fraction of the gridded region.
This process is repeated until enough grid points are within the image to 
calculate its area to better than 1\% accuracy.  The magnification is
then the ratio between this area and the original area of the unlensed
source.

If the substructure is massive ($\simgt 10^7\msun$) it can change the
position of the images enough that the contribution of the host lens to
the image magnification can change.  In addition, a single one of these
large substructures can affect the magnification of more
than one image.  For these reasons we simulate the whole lens (host lens
plus substructure) at once.  In this way we can measure the positions of
the images which become significant constraints on the substructure
when its mass is large.

CDM simulations predict that most of the substructure at a projected
galactic radius of a few kpc will be at a three dimensional radius of a
few tens of kpc \markcite{astro-ph/0304292}(Zentner \& Bullock 2003).  The
substructure will then be distributed fairly uniformly on the scale of the
host's Einstein radius. 
For this reason we do simulations with a constant number density of
substructures.  We do not characterize the simulations by the mass
fraction in substructure because this can vary by a factor of several
between the different images owing to the differences in density at
small radii.  The substructures are modeled as singular isothermal
spheres (SISs) tidally truncated as
if they are at a distance of 30~kpc from the center of an isothermal
galaxy.  The lensing is not critically dependent on the details of this
truncation.  All the substructures are of the same mass.  In a more
theoretically motivated model there would be a distribution of
substructure masses, but in this paper we are seeking to put empirical
constraints on the substructure mass scale.

\subsubsection{radial profile degeneracy}
\label{sec:degen}

There is a degeneracy between the level of substructure and the radial
profile of the host lens as discussed in \markcite{Metcalf2003}Metcalf (2003).  The effect
of changing the radial profile is investigated by adjusting the host
lens model.  We start with an elliptical lens model with an
$\kappa(\rho)\propto \rho^{-1}$ (where $\rho$ in coordinates that are aligned
with the axis of the lens is $\rho^2=x_1^2+x_2^2/(1-\epsilon)^2$) profile
that fits the image positions very well.  This is known as a singular
isothermal ellipsoid (SIE) lens model.  We then add a uniform surface
density, $\sigma$, to the lens and adjust the normalization of the
ellipsoidal component so that the image positions do not change.  This
is always possible and has no effect on the magnification ratios, but it
does change the total magnification such that flatter profiles have
higher magnifications and it changes the time delays such that flatter
profiles have shorter time delays.  The total magnification is
unobservable and the time delay has not been measured for Q2237+0305.
These degenerate models are labeled by their effective logarithmic
slope, $n_{\rm eff}$, which is related to the the added surface density
by $n_{\rm eff}=(\sigma-1)$. For further discussion see \markcite{Metcalf2003}Metcalf (2003).

For each choice of substructure properties and host lens model we run
simulations with three different source sizes -- 10~pc which is
appropriate for the radio and mid-IR emission regions and for the NLR we
use 200~pc and 400~pc for all.  This size should be viewed as an
effective size for the purposes of lensing not the full extent of the
NLR which must be larger.

\subsubsection{results}
\label{sec:results}

\begin{figure}[t]
\centering\epsfig{figure=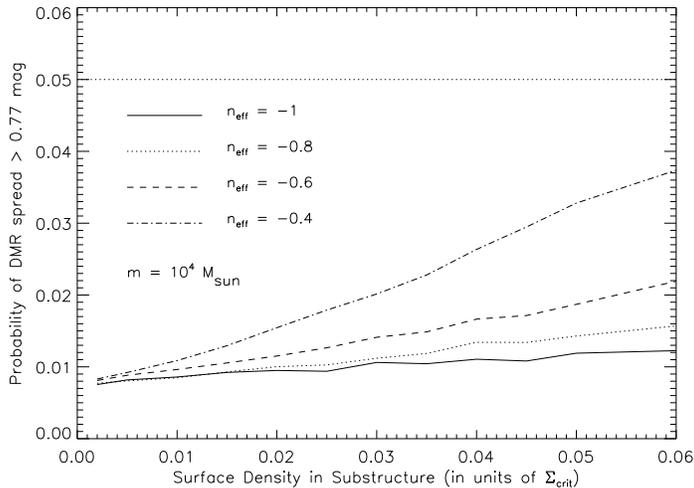,height=2.75in}
\caption[]{\footnotesize The probability that the spread in the DMRs,
  ${\mathcal S}$, will be greater than the observed 0.77~mag,
  $P({\mathcal S} \geq 0.77\,|\,\kappa_{\rm sub},n_{\rm eff},m)$, with
  substructure mass $m=10^4\msun$.  The different curves are for host
  lens models with different radial profiles (see
  Section~\ref{sec:degen}).  The horizontal dotted line marks the 5\%
  level for reference.  To convert the surface densities into mass
  fractions in substructure see Table~\ref{table2}.}
\label{fig:prob_m4}
\end{figure}

\begin{figure}[t]
\centering\epsfig{figure=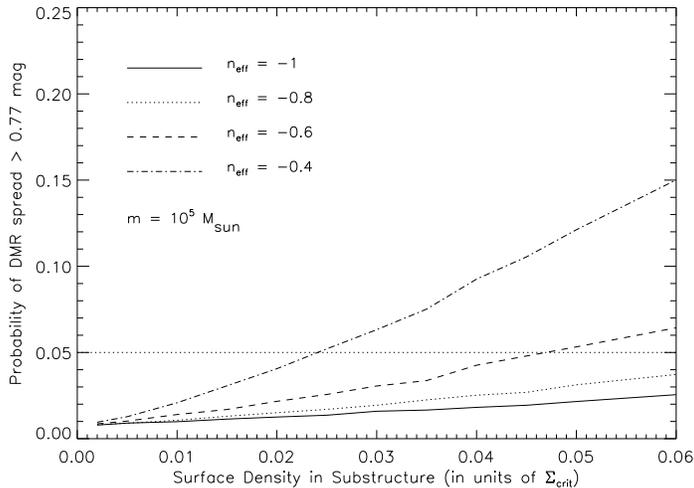,height=2.75in}
\caption[]{\footnotesize Same as in Figure~\ref{fig:prob_m4} only with
  substructure mass $m=10^5\msun$. Note that the scale is different and
  the probabilities are generally larger.}
\label{fig:prob_m5}
\end{figure}

\begin{figure}[t]
\centering\epsfig{figure=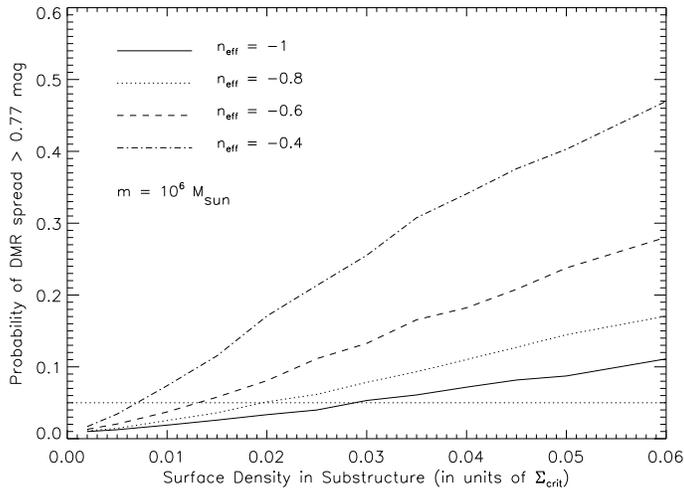,height=2.75in}
\caption[]{\footnotesize Same as in Figure~\ref{fig:prob_m5} only with
  substructure mass $m=10^6\msun$.  }
\label{fig:prob_m6}
\end{figure}

\begin{figure}[t]
\centering\epsfig{figure=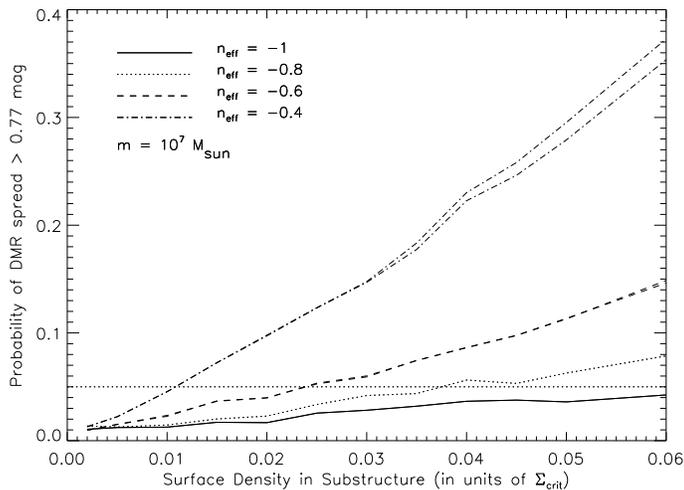,height=2.75in}
\caption[]{\footnotesize Same as in Figure~\ref{fig:prob_m6} only the
  substructure mass is $10^7\msun$ and two limits on the shift of the
  image positions are shown.  For each $n_{\rm eff}$ the lowest curve is for a
  $0\farcs 2$ limit and the highest is for no limit.  For the $n_{\rm
  eff}=-1$, $-0.8$ and $-0.6$ the $0\farcs 2$ 
  limit is indistinguishable from no limit.  Note that scale in this
  plot is not the same as in Figure~\ref{fig:prob_m6}.}
\label{fig:prob_m7}
\end{figure}

\begin{figure}[t]
\centering\epsfig{figure=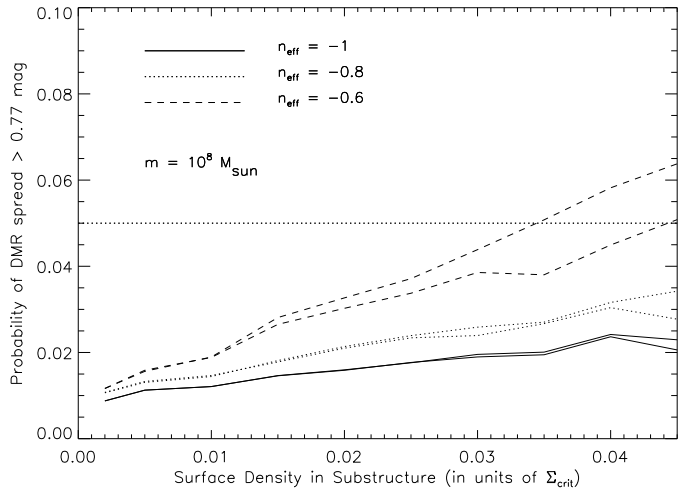,height=2.75in}
\caption[]{\footnotesize Same as in Figure~\ref{fig:prob_m7} except the
  substructure mass is $10^8\msun$. None of the models reached above 0.065
  probability that the DMRs spread will be greater than 0.77~mag.}
\label{fig:prob_m8}
\end{figure}

\begin{table}[t]
\begin{center}
\begin{tabular}{clrrrrr}
Image  & $n_{\rm eff}=$ & $-1.0$ & $-0.8$ & $-0.6$ & $-0.4$ & $-0.2$ \\
\hline
A &$\kappa=$& 0.4 & 0.5 & 0.6 & 0.8 & 0.9\\
B &$\kappa=$& 0.4 & 0.5 & 0.6 & 0.8 & 0.9\\
C &$\kappa=$& 0.7 & 0.8 & 0.8 & 0.9 & 0.9\\
D &$\kappa=$& 0.6 & 0.7 & 0.8 & 0.8 & 0.9\\
\end{tabular}
\caption[dw]{\footnotesize The surface densities at each of the images in
  units of $\Sigma_{\rm crit}$ ($= 9.8\times 10^3\msun\pc^{-2}$ in the
  $H_o= 70 \kms\mpc^{-1}$, $\Omega_\Lambda=0.7$, $\Omega_{\rm matter}=0.3$
  cosmology.  These values differ from those in the lens model of
  \markcite{1998MNRAS.295..488S}{Schmidt}, {Webster}, \&  {Lewis} (1998) at the 10\% level.  The $n_{\rm eff}=-1$
  model is the singular isothermal ellipsoid lens model. }
\label{table2}
\end{center}
\end{table}

\begin{table}[t]
\begin{center}
\begin{tabular}{crrrrrrrr}
\small
 $\ell_{\rm NLR}$ (pc)  & $200$ & $200$ & $200$ & $200$ & $400$ & $400$ & $400$ & $400$\\
 $n_{\rm eff}$ & $-1.0$ & $-0.8$ & $-0.6$ & $-0.4$ & $-1.0$ & $-0.8$ & $-0.6$ & $-0.4$ \\
\hline
$10^4\msun$ & $>0.06$ & $>0.06$ & $>0.06$ & $>0.06$ & $>0.06$ & $>0.06$
 & $>0.06$ & $>0.06$  \\
$10^5\msun$ & $>0.06$ & $>0.06$ & $0.047$ & 0.024 & $>0.06$ & $>0.06$ &
 0.047 & 0.024\\
$10^6\msun$ & 0.029 & 0.020 & 0.013 & 0.007 & 0.034 & 0.023 & 0.015 & 0.008\\
$10^7\msun$ & $>0.06$ & 0.037 & 0.024 & 0.011 & 0.053 & 0.030 & 0.017 & 0.007 \\
$10^8\msun$ & -- & -- & 0.034 & & -- & 0.035 & 0.019 \\
\end{tabular}
\caption[dw]{\footnotesize Minimum substructure surface densities
  required to have a better than 5\% chance of observing a DMR spread as
  large as the one observed.  The cases marked ``$>0.06$'' did not reach
  this limit within the range of surface densities tested, those marked
  ``--'' appear as though they cannot reach this limit for any
  $\kappa_{\rm sub}$ and those that are blank where not tested because
  the images tended to merge in too many cases. }
\label{table_kappa}
\end{center}
\end{table}

\begin{figure}[t]
\centering\epsfig{figure=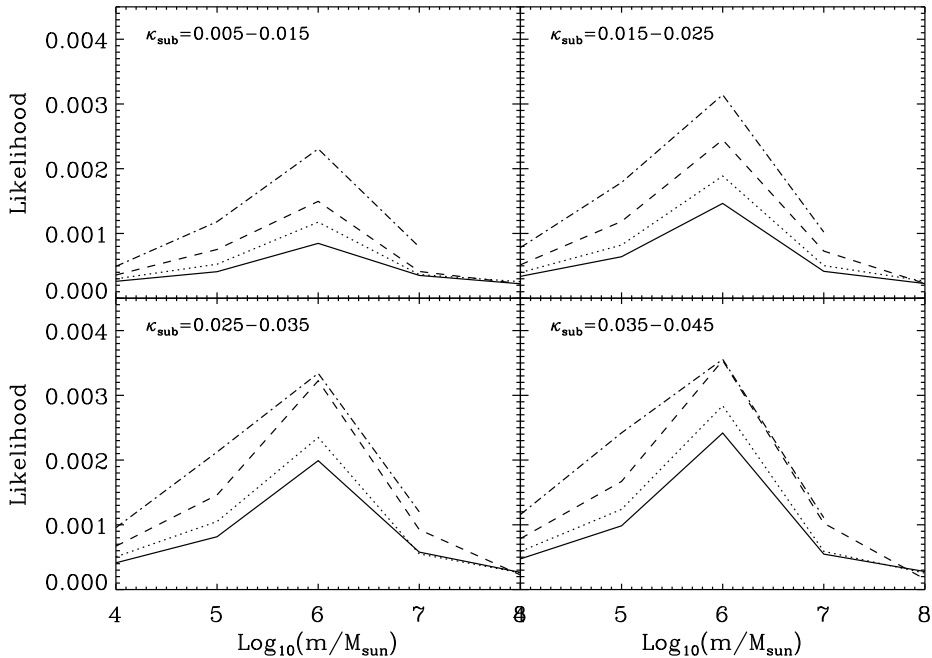,height=3.5in}
\caption[]{\footnotesize These are the likelihood functions for four ranges
  of substructure surface density as marked in each panel.  The curves are
  marked in the same way as in Figures~\ref{fig:prob_m4} through
  \ref{fig:prob_m8} -- solid: $n_{\rm eff}=-1$, dotted: $n_{\rm eff}=-0.8$,
  dashed: $n_{\rm eff}=-0.6$ and dot-dashed: $n_{\rm eff}=-0.4$.  The
  limit in the acceptable shift in the image positions is $0\farcs 2$.
  The normalization is arbitrary, but all the models are normalized in
  the same way so that they can be compared between panels.  A
  substructure mass of $\sim 10^6\msun$ is clearly favored in all cases.}
\label{fig:likelihood}
\end{figure}

\begin{figure}[t]
\centering\epsfig{figure=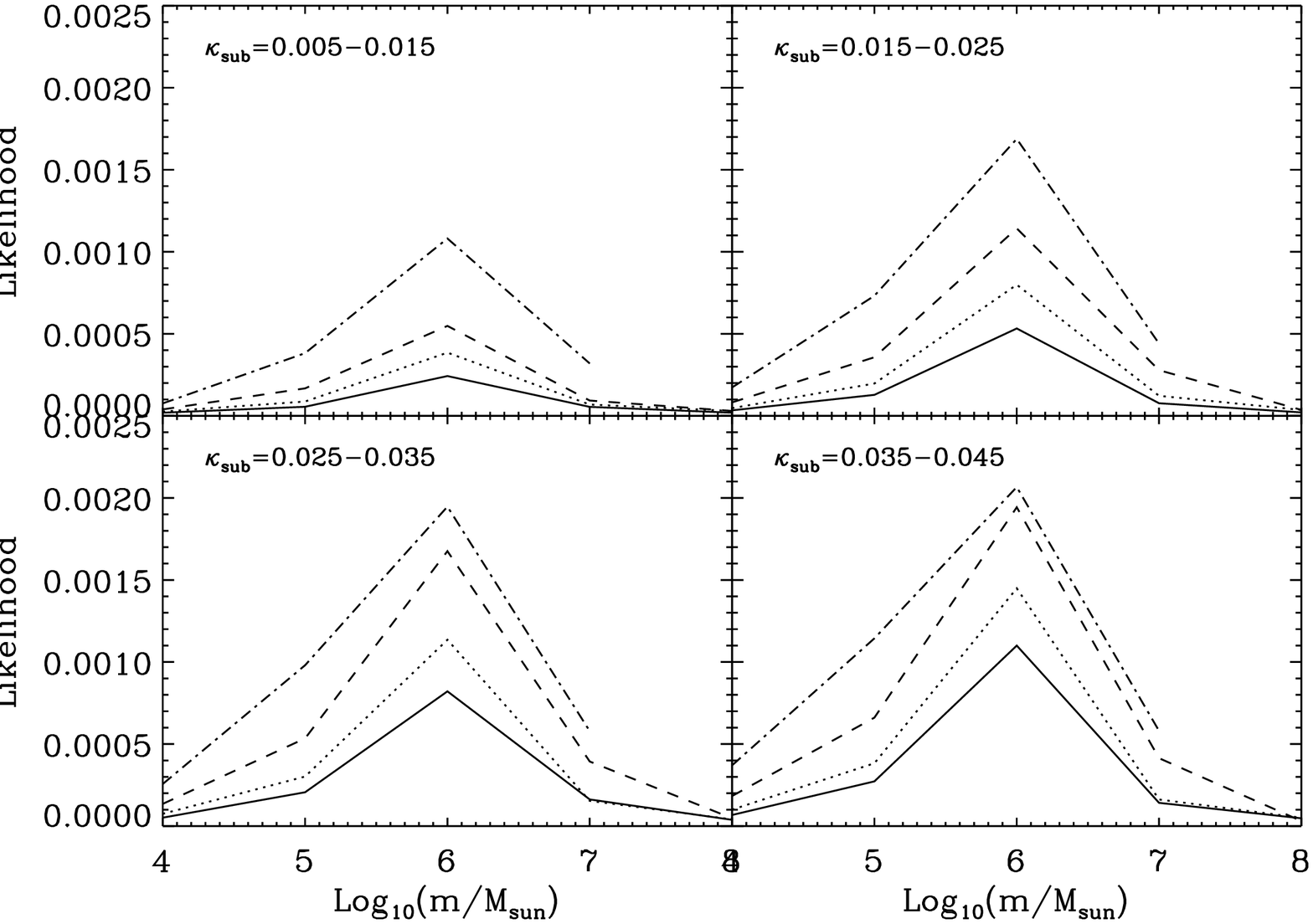,height=3.5in}
\caption[]{\footnotesize The same as in Figure~\ref{fig:likelihood} only
the errors for the NLR magnification ratios reported in
Table~\ref{table} are used instead of doubling them to account for
unknown systematic errors.  The normalization is arbitrary  and different
from that in Figure~\ref{fig:likelihood}.} 
\label{fig:likelihood2}
\end{figure}

We find that the data are best suited to putting a lower limit on the
surface density of substructure as a function of the substructure mass
and the host lens profile; this is the most robust and conservative way
to treat the data.  To quantify this we use the {\it spread} statistic.
We define the spread, ${\mathcal S}$, as the difference between the
largest DMR and the smallest, including the normalization image's DMR
($\Delta=0$).  This statistic is clearly independent of which image is
used to normalize the DMRs.  In the absence of substructure and noise
${\mathcal S}=0$.  As shown in Section~\ref{sec:comp-with-radio} the
measured spread is ${\mathcal S}=0.77\mag$.  We calculate the probability
of getting an observed spread as large as this or larger given the level
of substructure and the noise, $P({\mathcal S}\geq 0.77)$.  This is done by
Monte Carlo, using the simulations to predict the flux of each image and
then adding random noise.  At least 2,000 lensing simulations are
calculated for each choice of substructure parameters and for each
simulation 500 realizations of the noise are made.  The results do not
change when these numbers are increased.  To be conservative we use
twice the NLR flux errors derived from the data and reported in Table~\ref{table}.  

Figures~\ref{fig:prob_m4} through \ref{fig:prob_m8} show the probability
$P({\mathcal S}\geq 0.77)$ for different substructure masses.
Figure~\ref{fig:prob_m4} is for a mass of $10^4\msun$.  In this case the
substructures are not large enough to change the magnification of the
NLR and only rarely can significantly change the magnification of the
10~pc source representing the radio and mid-IR sources.  The probability
is correspondingly low with none of the models going above 5\%
probability for $\kappa_{\rm sub}<0.06$.  We 
consider this low probability unacceptable thus ruling out
$m=10^4\msun$ substructure with $\kappa_{\rm sub}<0.06$ as an explanation.
To translate these values into mass fractions
Table~\ref{table2} gives the surface densities at each image in the
different models.  

 Figure~\ref{fig:prob_m5} shows the probabilities for mass $10^5\msun$.
 The probability is larger here than for $10^4\msun$, but still not
 significant unless the host halo is flatter than an SIE.   The $n_{\rm
 eff}=-1$ model, the most common assumption in lensing, is ruled out for
 the range of $\kappa_{\rm sub}$ tested.  Our maximum surface density of
 $\kappa_{\rm sub}=0.06$ is still very large compared to 
 the some CDM predictions of less than 0.1\% at this mass scale \markcite{astro-ph/0304292}(Zentner \& Bullock 2003).
 The $n_{\rm eff}=-0.4$ case reaches our thresh-hold of $5\%$ at $\kappa_{\rm sub}=0.047$
 making this the lower limit in this case.

 A mass of $10^6\msun$ is investigated in Figure~\ref{fig:prob_m6}.
In this case for substructure surface densities $\kappa_{\rm sub} >
0.029$ the observed DMR 
discrepancy occurs more than 5\% of the time in all the host models
tested.  At the 10\% level $\kappa_{\rm sub} > 0.058$ in the most common
$n_{\rm eff}=-1$ model. For flatter models this lower limit is reduced
significantly.  These are more reasonable mass fractions since $\kappa_{\rm
  sub} =0.02$ corresponds to $f_{\rm sub}\simeq 0.05$ at image~A in the
$n_{\rm eff}=-1$ model and less in the flatter models.  However, at this
 mass scale this is still rather large compared to the current CDM predictions.

Figure~\ref{fig:prob_m7} shows $P({\mathcal S}\geq 0.77)$ for substructure of mass
$10^7\msun$.  In this case the substructure is big enough to occasionally
shift the image positions by a few tenths of an arcsec.  Lens models for
Q2237+0305 using the observed distribution of stars have been very
successful at reproducing the image positions 
\markcite{2002MNRAS.334..621T,1998MNRAS.295..488S}({Trott} \& {Webster} 2002; {Schmidt} {et~al.} 1998).  For this reason any
substructure cannot change the image 
position by very much.  In Figure~\ref{fig:prob_m7} limits on the
allowed shift in image positions are imposed to express this additional
constraint.  We adopt a conservative limit of $0\farcs 2$ as a test.
Results with and without the image shift constraint are plotted in
Figure~\ref{fig:prob_m7} and it is seen that this extra constraint does
not change the lower limits on $\kappa_{\rm sub}$.  The most striking
result displayed in  
Figure~\ref{fig:prob_m7} is that the $n_{\rm eff}=-1$ is unlikely to produce the
observations for any substructure surface density that is reasonable
from the prospective of current CDM model predictions ($\kappa_{\rm sub}=0.06$ would be
15\% of the mass at image~A).  
In fact, $P({\mathcal S}\geq 0.77)$ does not go
above 15\% within the range of $\kappa_{\rm sub}$ investigated unless $n_{\rm
  eff} \geq -0.4$.
In some cases we can put an upper limit on $\kappa_{\rm sub}$ and they
are reported in Table~\ref{table_kappa}.

Figure~\ref{fig:prob_m8} shows $P({\mathcal S}\geq 0.77)$ substructure of mass
$10^8\msun$.  Here all the models are ruled out at the 90\% level and
the only way to get above the 95\% level is to abandon a limitation on
the shift of the image position which we think is not realistic.  High
mass substructure is not compatible with the observed DMRs because it is
not effective in changing the small and large scale magnification ratios
independently - when it magnifies (or demagnifies) the radio and mid-IR
images it also magnifies (or demagnifies) the NLR image.  This is
further conformation that anomalous DMRs must be the results of small
scale structure and not a large scale distortion of the host lens.  In
addition, the $10^8\msun$ substructure tends to displace the images on
the scale of several tenths of an arcsec although this mass range is
ruled out even without this constraint.

We have done these calculations with source sizes of $200\pc$ and
$400\pc$ representing the NLR.  Changing the size makes only small
changes to $P({\mathcal S}\geq 0.77)$ at large mass and large $n_{\rm eff}$.
These changes do not significantly change the conclusions.  For
Figures~\ref{fig:prob_m4} through \ref{fig:prob_m8} the NLR source size
is set to $200\pc$.  The lower limits on $\kappa_{\rm sub}$ for all the 
models tested are shown in table~\ref{table_kappa}.

The error estimates that have been derived for our observations
and those published with the mid--IR and radio observations are crucial
for the estimation of the surface density bound.  If the errors are underestimated the
$\kappa_{\rm sub}$ lower bound should be lower.  For example, if we
increase the radio errors of \markcite{1996AJ....112..897F}{Falco} {et~al.} (1996) by half the
$\kappa_{\rm sub}$ lower limit in the $n_{\rm eff}=-1$, $m=10^6\msun$
model goes from 0.029 to 0.021.  Increasing the mid--IR errors
\markcite{2000ApJ...545..657A}({Agol} {et~al.} 2000) in this way changes this limit to 0.016.  On
the other hand if the empirically derived errors on our NLR measurements
are used instead of doubling them the limit is increased to 0.055.

To squeeze more out of the data we can also calculate likelihood
functions for the substructure parameters although this is somewhat less
robust then the $P({\mathcal S}\geq 0.77)$ analysis done above.  The
likelihood function for set of parameters $\{\kappa_{\rm sub},n_{\rm
  eff},m\}$ is given by 
\begin{eqnarray}\label{likelihood}
{\mathcal L}(\kappa_{\rm sub},n_{\rm eff},m) = 
\int d\{\dmr\}~ P_{\rm sub}\left(\{\dmr\},\{{\bf x}\}\,|\,\kappa_{\rm
  sub},n_{\rm   eff},m\right)
\\ \times P_{ob}\left(\{\dmr_{ob}\},\{{\bf x}_{ob}\}\,|\,\{\dmr\},\{{\bf x}\}\right) \nonumber
\end{eqnarray}
where $P_{ob}$ expresses the instrumental uncertainties which is taken
to be a normal distribution.  Uniform prior distributions have been
assumed for the parameters.  Again the integral is done by Monte Carlo.

Figure~\ref{fig:likelihood} shows the likelihoods as functions of
substructure mass in several ranges of substructure surface density.
Plotted this way it is clear that the data favors a substructure mass
of $\sim 10^6\msun$ for fixed $\kappa_{\rm sub}$.  Here the errors in
the NLR magnification ratios have been set to twice those in
Table~\ref{table} to be conservative.  The likelihood functions with the
measured statistical errors are shown in Figure~\ref{fig:likelihood2}.
The shape of the likelihood function is very similar, but the extreme
masses are more strongly ruled out.

In summary, the data puts a lower limit on the amount of substructure
that is present in Q2237+0305.  Table~\ref{table_kappa} gives these
limits for the ranges of parameters tested.  Given the present CDM
simulations these limits seems uncomfortably large.
With the expected substructure level of $<1\%$ of the surface density
in subclumps of mass less than $10^8\msun$ the host lens profile as
steep as $\kappa(\rho)\propto \rho^{-1}$ is unacceptable.  We also find that a
substructure mass scale of $\sim 10^6\msun$ is most effective at
changing the observed DMRs.  Substructure of mass $10^4\msun$ is too
small and $10^8\msun$ is too large.  The details of the lower mass limit
is dependent on our assumption that the radio and mid-IR emission
regions are $10\pc$ in radius, however to the precision we are now
concerned with this should not be a major factor.

It has been assumed here that the substructure is within or near the
lens galaxy.  It is possible that the responsible objects are in
intergalactic space either in front of or behind the lens.  Within the
CDM model this is less probable, but not improbable.  The above constraints
can be crudely converted to a redshift $z$ by scaling the substructure
mass and surface density by a factor of 
$\Sigma_{\rm crit}(z_s,z_l)/\Sigma_{\rm
  crit}(z_s,z)=\frac{D(z)\left\{D_s-D(z)\right\}}{D_l\{D_s-D_l\}}$.  This
factor reaches a minimum of 0.6 at $z=1.1$.

\subsubsection{The contribution of known objects}

There are known objects in our galaxies and others which could
conceivably be causing the observed anomalies in the DMRs.  One
possibility is globular clusters.  However, the total mass in
globular clusters around the Milky Way is only $\sim 10^7-10^8\msun$ so their
surface density is two order of magnitude below the lower limits
reported here.

Another contribution is from Giant Molecular Clouds (GMCs) of mass $\sim
10^6\msun$ which are known to contain most of the H$_2$ gas mass in the
Milky Way.  The surface density of H$_2$ at galactic radius $\simeq
1\kpc$ (where the images in Q2237+0503 form) is only a few
$\msun\pc^{-2}$, well below our lower limits \markcite{blitz97}(Blitz 1997).  The H$_2$
density in M31 no more than $\simeq 5 \msun\pc^{-2}$ anywhere
\markcite{1999A&A...351.1087L}({Loinard} {et~al.} 1999).  However, in some galaxies the peak H$_2$ density
does exceed our lower limit \markcite{2003ApJS..145..259H}({Helfer} {et~al.} 2003).  The H$_2$
density is truncated at an inner radius of several kpc.  This is
believed to be caused by the galactic bar.  The same phenomena is seen
in other barred spirals.  In Q2237+0503 the lensing is almost entirely
caused by the bulge and bar of the galaxy through which the images pass
\markcite{2002MNRAS.334..621T}({Trott} \& {Webster} 2002).  For this reason we expect the density of
GMCs to be well below our limit.  In addition, GMCs are not very
centrally concentrated, they are generally held up by turbulent pressure
and have many, much smaller mass, condensations sprinkled throughout
them.  For this reason GMCs make inefficient lenses as compared to the
SIS model used in Section~\ref{sec:simulations} through
\ref{sec:results}.  The GMCs would need more density by a factor of a
few to reproduce reproduce the lensing observations.  

Another argument for a small surface density in GMCs is the extinction.  
By adopting an upper limit on the absolute luminosity of the QSO of
$M_B > -29$ and an upper limit of 20 for the lens magnification (the SIE+shear
model predicts 16) an upper limit for the extinction of $A_V < 2.7\mag$
can be derived \markcite{1999ApJ...523..617F}({Falco} {et~al.} 1999).  The extinction in the Milky
Way is roughly related to the hydrogen column density through
$N(H\,{\scriptsize I})+2N(H_2)\simeq 13\, A_V \msun \pc^{-2}$
\markcite{1977ApJ...216..291S}({Savage} {et~al.} 1977).  Conservatively assuming all the hydrogen
is in molecular form this gives an upper limit of $N(H_2)< 18\msun
\pc^{-2}$ which is well below our lower limit on
substructure. \markcite{1999ApJ...523..617F}{Falco} {et~al.} (1999) have argued that the above
relation between hydrogen column density and extinction does not hold
for some of the lens galaxies they studied in the sense that it
over estimates the column density which makes our upper limit even more
conservative.

For the above reasons we do not think GMCs are a likely explanation for the
DMR anomalies, but they are probably the largest potential
``background'' and we plan to investigate their lensing properties more
thoroughly in the future.  For lenses with elliptical galaxies -- the
majority of QSO lenses -- GMCs are much less likely to be a source of
lensing substructure.  Analyzing more lens systems can definitively rule
of GMCs.

\section{Discussion}
\label{sec:discussion}

We have demonstrated the spectroscopic gravitational lensing method and
shown the importance of having high quality data in a large range of
wavelength.  Using the CIRPASS instrument the magnification ratios of
the QSO's narrow lines region and broad line region were measured.  It
was found that these do not agree with each other or with the published
radio and mid-IR magnification ratios.  The disagreement between the BLR
magnification ratios and the radio/mid-IR magnification ratios is
interpreted as microlensing of the 
BLR.  This implies that the mid-IR emission region is larger than the
BLR and the BLR is $\simlt 0.1\pc$ in size.  We find a small difference in
the shape of the H$\beta$ line in image~A when compared to the other
images.  This difference is too small and symmetric to be considered
evidence for rotation or large scale infall in the H$\beta$ emission
region.  The disagreement between the radio/mid-IR ratios and the NLR
ratios is interpreted as a signature of substructure.  Lensing
simulations are performed to obtain a lower limit on the amount and mass
of substructure that is required to cause this discrepancy.  The
substructure content is degenerate with the radial profile of the host
lens, but if the expectations of the CDM model are taken into account
certain radial profiles and substructure surface densities can be ruled
out.  The substructure mass scale as large as $10^8\msun$ and
$10^7\msun$ is disfavored while $10^4\msun$ is too small given our
assumptions about the radio/mid-IR source size.  The standard elliptical
singular isothermal lens mass profile is disfavored in comparison to a more
flattened profile with the same substructure content.

Although the anomalous differential magnification ratios in Q2237+0305
are very unlikely to be caused by a substructure as large as $10^8\msun$
this does not mean that the mass function of substructures does not
extend up to this mass scale.  Large mass substructures could dominate
the total mass in substructure and yet be very few in number so that
one would expect many lenses to be unaffected by them; in other words
the shot noise from lens to lens is expected to be large at high
masses.  If the responsible object is intergalactic and unassociated
with the lens galaxy the mass bound is different, but in this case the
upper mass bound is more likely to go down then up because the lens
galaxy is unusually close to us and not near the minimum of $\Sigma_{\rm
  crit}(z_s,z)$. 

It should be mentioned that flattening the host lens' radial mass
distribution generally reduces the Hubble parameter as measured by the
time delays.  The X-ray measurement of the time delay
($2.7^{+0.5}_{-0.9}$~hours between image~A and B) is tentative
\markcite{2003ApJ...589..100D}({Dai} {et~al.} 2003), but it does indicate that Q2237+0305 has a
near $r^{-1}$ surface density profile if the HST Key Project's value for
the Hubble parameter is used ($72\pm 8 \kms\mpc^{-1}$;
\markcite{2001ApJ...553...47F}{Freedman} {et~al.} 2001).  The fiducial SIE+shear model used here gives a
time delay of 3.3~hours and the delay is proportional to $n_{\rm eff}^{-1}$ as
defined in section~\ref{sec:degen}.  In general the measured time delays
in other lenses lead to a conclusion that either the Hubble
parameter is smaller than the above value or the profile is steeper than
$r^{-1}$.  \markcite{2003ApJ...583...49K}{Kochanek} (2003) has highlighted this as a problem
for the CDM model.  This is an unresolved and intriguing problem which
becomes even more so with our results since the steeper the radial
profile the more substructure is needed.  It could be that in looking
for substructure to confirm CDM we have found too much.

Clearly more lens systems need to be studied in this way to put stronger
constraints on dark matter substructure and eliminate systematics.  The
prospects for improvement are good in this respect since Q2237+0305 is
not an ideal system for this work -- the \oiii lines are in a region of
high atmospheric absorption and because of the proximity of the lens
galaxy microlensing is enhanced.  The situation can be greatly improved
with just a few more systems observed in narrow lines and the mid-IR
and/or radio. 
One of the most important systematic uncertainties we encountered is how
to construct the apertures around each of the images for measuring the
magnification ratios.  The improved spatial resolution that would come with
using a space-based observations (or with adaptive optics-enhanced
spectroscopy from the ground) would be of great benefit in reducing this
uncertainty.

\appendix
\section{Contribution of Wavelength Dependent Magnification}
\label{sec:contr-wavel-depend}

\begin{figure}[t]
\centering\epsfig{figure=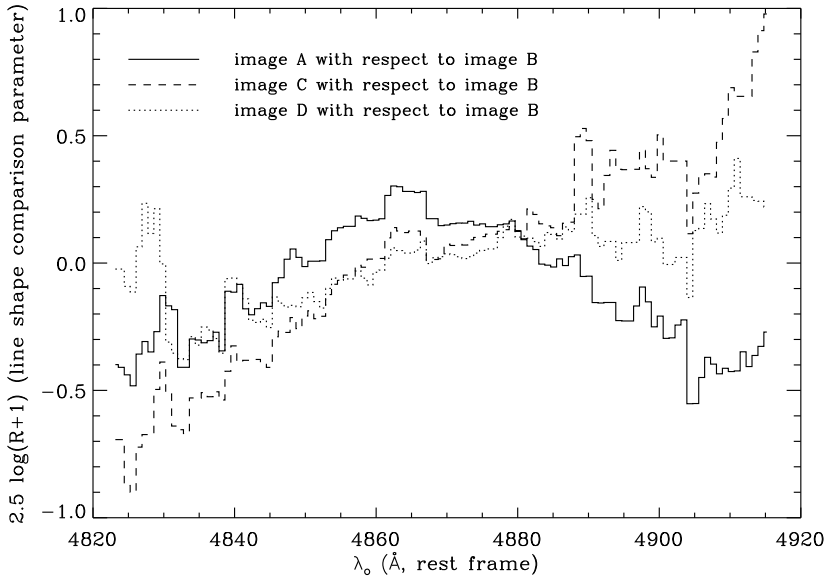,height=3.25in}
\caption[]{\footnotesize The line profile distortion parameter,
  equation~\ref{Delta_F}, between image~A and the other images.  The
  central peak is 0.3~mag which indicates that the wavelength
  independent contribution to the NLR/BLR differential magnification
  ratio is at minimum 0.3~mag.
}\label{fig:line_shape_R}
\end{figure}

It is possible that the magnification of the BLR is not uniform as would
probably be the case if microlensing is
important and the BLR size is of order the Einstein radius of the stars
in the lens galaxy.  This possibility can be tested if the BLR has a
coherent 
large scale velocity structure as in the case of a rotating disk or
spherical infall.  In this case a small scale variation in the
magnification would distort the the broad line in one or more of the
images.  This effect was proposed by \markcite{1988ApJ...335..593N}{Nemiroff} (1988).  The contribution of
such a wavelength dependent magnification to the differential
magnification ratio can be assessed by comparing the line profiles from
the two images.

If the wavelength of the $i$th pixel of a spectrum is $\lambda_i$ then
the flux in this pixel is $F(\lambda_i)=\mu(\lambda_i)f(\lambda_i)$
where $f(\lambda_i)$ is the flux before magnification and
$\mu(\lambda_i)$ is the magnification at that wavelength.  The total
flux from the BLR is $F=\sum_i F(\lambda_i)$.  We will assume that
one of the BLR images is uniformly magnified, i.e. the magnification is not a
function of wavelength.  To be specific we will label this image B
although it could be any image or an average of some of the images.  For
this image $F^B(\lambda_i)/F^B(\lambda_o)=f(\lambda_i)/f(\lambda_o)$
where $\lambda_o$ is a reference pixel that can be varied.  Because the
pre-lensed line profile, $f(\lambda_i)$, is the 
same for all the images we can calculate the change to the total flux in
image~A caused by a frequency dependent magnification:
\begin{equation}\label{Delta_F}
R(\lambda_o) \equiv \frac{F^A-F^A_o}{F^A_o}=\frac{F^B(\lambda_o)F^A}{F^A(\lambda_o)F^B} - 1
\end{equation}
where $F^A_o$ is what the total flux would be if the magnification were
uniform and equal to $\mu^A(\lambda_o)$.  Another way to think of this
quantity is to note that if you wanted to correct the observed
differential magnification ratio (with the sign convention used in this
paper) to what it would be if the BLR magnification
ratio where uniformly $\mu^B/\mu^A(\lambda_o)$ then one would subtract
$2.5\log[R(\lambda_o)+1]$ magnitudes.
After subtracting the continuum and masking any sky lines 
equation~(\ref{Delta_F}) can be applied directly to the data to determine
if a wavelength dependent component of the magnification is causing a
significant change in the line flux.  Figure~\ref{fig:line_shape_R}
shows this for our data.

Microlensing is not the only thing that could affect the line profiles.
Within the broad line there is also a contribution from the NLR.
This NLR contamination will reduce the magnitude of the differential
magnification ratio, diluting any signal from substructure.  It will
also make images with relatively bright BLRs have a wider line profiles
as is observed in our data (Figure~\ref{fig:Hbeta_lines}). 
If we use a simple model where there is a distinct NLR and BLR in
H$\beta$ (which is probably not the case) and we denote their
magnifications as $\mu_{\rm NL}$ and $\mu_{\rm BL}$ respectively then
the expected shape parameter will be
\begin{equation}\label{R_NL}
R(\lambda_o) = 
\frac{(\omega(\lambda_o)-\omega)(r^A-r^B)}{(\omega(\lambda_o)+r^A)(\omega+r^B)}
\end{equation}
where $\omega=f_{\rm NL}/f_{\rm BL}$ and $r=\mu_{\rm BL}/\mu_{\rm NL}$.
Note that if the NLR and BLR components of the line are both symmetric then
this effect will not make the line asymmetric. More generally, since the
NL is narrow by definition the line profile should only be modified in the
central region.  

It is also interesting to note that in the limit that
$\omega\rightarrow 0$ while the central wavelength is still dominated by
the NL ($\omega(\lambda_{\mbox{H}\beta})\rightarrow \infty$)
$R(\lambda_{\mbox{H}\beta})=\mbox{DMR}-1$.  With our data this would
give an estimate of $\mbox{DMR}=0.3\mag$ between A and B~images - still well
above the noise.  This is a lower limit since as the dominance of the NL
at the central wavelength, $\omega(\lambda_{\mbox{H}\beta})$, is reduced
the implied DMR becomes larger.

\section*{Acknowledgments}
\footnotesize
 This paper is based on observations
obtained at the Gemini Observatory, which is operated by the Association
of Universities for Research in Astronomy, Inc. (AURA), under a
cooperative agreement with the U.S. National Science Foundation (NSF) on
behalf of the Gemini partnership: the Particle Physics and Astronomy
Research Council (UK), the NSF (USA), the National Research Council
(Canada), CONICYT (Chile), the Australian Research Council (Australia),
CNPq (Brazil) and CONICET (Argentina). We are grateful to Matt Mountain
for the Director's discretionary time to demonstrate the scientific
potential of integral field units (the PIs of this demonstration science
program are Andrew Bunker, Gerry Gilmore and Roger Davies). We thank the
Gemini Board and the Gemini Science Committee for the opportunity to
commission CIRPASS on the Gemini-South telescope as a visitor
instrument. We received excellent support from Gemini, and thank Phil
Puxley, Jean Ren\'{e}-Roy, Doug Simons, Bryan Miller, Tom Hayward,
Bernadette Rodgers, Gelys Trancho, Marie-Claire Hainaut-Rouelle and
James Turner. CIRPASS was built by the instrumentation group of the
 Institute of Astronomy (IoA) in Cambridge, UK. We warmly thank the Raymond
 and Beverly Sackler Foundation and PPARC for funding this
 project. Andrew Dean, Rob Sharp, Anamparambu Ramaprakash and Anthony
 Horton all assisted with the observations in Chile, and we are indebted
 to Dave King, Jim Pritchard \& Steve Medlen for contributing their
 instrument expertise. The optimal 
extraction software for this 3D fiber spectroscopy was written by Rachel
Johnson, Rob Sharp and Andrew Dean.  We would like to thank J. Primack
 for providing access to the UC Santa Cruz beowolf computer cluster.
RBM would like to thank the IoA for hosting him during part of the
 preparation of this paper and P. Natarajan for helpful discussion.  
LAM acknowledges support from the SIRTF
 Legacy Science Program provided through an award issued by Jet
 Propulsion Laboratory, California Institute of Technology, under NASA
 contract 1407.  Financial support for RBM was provided by
 NASA through Hubble Fellowship 
grant HF-01154.01-A awarded by the Space Telescope Science Institute,
which is operated by the Association of Universities for Research 
in Astronomy, Inc., for NASA, under contract NAS 5-26555

\end{document}